\documentclass[twocolumn]{aastex63}

\usepackage{amsmath}
\usepackage{booktabs}

\newcommand{\hddeux}{HD\,209458\,b}

\newcommand{\hdun}{HD\,189733\,b}
\newcommand{\hduns}{HD\,189733\,b }
\newcommand{\water}{H$_2$O}
\newcommand{\methane}{CH$_4$}
\newcommand{\diox}{CO$_2$}
\newcommand{\tp}{\ensuremath{T_{\rm P}}}
\newcommand{\kp}{$K_{\rm P}$}
\newcommand{\vsys}{$v_{\rm sys}$}

\newcommand{\vv}[1]{$v_{\rm #1}$}
\newcommand{\vmr}{\ensuremath{\log_{10}\text{VMR}[{\rm H}_2{\rm O}]}}
\newcommand{\cloud}{$P_{\rm cloud}$}
\newcommand{\logl}{$\ln \mathcal{L}$}
\newcommand{\kms}{km\,s$^{-1}$}

\shortauthors{Boucher et al.}

\begin{document}

\title{\textsc{Characterizing exoplanetary atmospheres at high resolution with SPIRou: \\ Detection of water on HD\,189733\,b}}
\correspondingauthor{Anne Boucher}
\email{anne.boucher.3@umontreal.ca}



\author[0000-0001-9427-1642]{Anne Boucher} 
\affiliation{Institut de Recherche sur les Exoplan\`etes, Universit\'e de Montr\'eal, D\'epartement de Physique, C.P. 6128 Succ. Centre-ville, Montr\'eal, QC H3C 3J7, Canada.}

\author[0000-0002-7786-0661]{Antoine Darveau-Bernier}
\affiliation{Institut de Recherche sur les Exoplan\`etes, Universit\'e de Montr\'eal, D\'epartement de Physique, C.P. 6128 Succ. Centre-ville, Montr\'eal, QC H3C 3J7, Canada.}

\author[0000-0002-8573-805X]{Stefan Pelletier}
\affiliation{Institut de Recherche sur les Exoplan\`etes, Universit\'e de Montr\'eal, D\'epartement de Physique, C.P. 6128 Succ. Centre-ville, Montr\'eal, QC H3C 3J7, Canada.}

\author[0000-0002-6780-4252]{David Lafreni\`ere}
\affiliation{Institut de Recherche sur les Exoplan\`etes, Universit\'e de Montr\'eal, D\'epartement de Physique, C.P. 6128 Succ. Centre-ville, Montr\'eal, QC H3C 3J7, Canada.}

\author[0000-0003-3506-5667]{\'Etienne Artigau}
\affiliation{Institut de Recherche sur les Exoplan\`etes, Universit\'e de Montr\'eal, D\'epartement de Physique, C.P. 6128 Succ. Centre-ville, Montr\'eal, QC H3C 3J7, Canada.}

\author[0000-0003-4166-4121]{Neil~J.~Cook}
\affiliation{Institut de Recherche sur les Exoplan\`etes, Universit\'e de Montr\'eal, D\'epartement de Physique, C.P. 6128 Succ. Centre-ville, Montr\'eal, QC H3C 3J7, Canada.}

\author[0000-0002-1199-9759]{Romain Allart}
\altaffiliation{Trottier Fellow}
\affiliation{Institut de Recherche sur les Exoplan\`etes, Universit\'e de Montr\'eal, D\'epartement de Physique, C.P. 6128 Succ. Centre-ville, Montr\'eal, QC H3C 3J7, Canada.}

\author[0000-0002-3328-1203]{Michael Radica}
\affiliation{Institut de Recherche sur les Exoplan\`etes, Universit\'e de Montr\'eal, D\'epartement de Physique, C.P. 6128 Succ. Centre-ville, Montr\'eal, QC H3C 3J7, Canada.}

\author[0000-0001-5485-4675]{Ren\'e Doyon}
\affiliation{Institut de Recherche sur les Exoplan\`etes, Universit\'e de Montr\'eal, D\'epartement de Physique, C.P. 6128 Succ. Centre-ville, Montr\'eal, QC H3C 3J7, Canada.}

\author[0000-0001-5578-1498]{Bj\"orn Benneke}
\affiliation{Institut de Recherche sur les Exoplan\`etes, Universit\'e de Montr\'eal, D\'epartement de Physique, C.P. 6128 Succ. Centre-ville, Montr\'eal, QC H3C 3J7, Canada.}

\author[0000-0002-0111-1234]{Luc Arnold}
\affiliation{CFHT Corporation; 65-1238 Mamalahoa Hwy; Kamuela, Hawaii 96743; USA}

\author[0000-0001-9003-8894]{Xavier Bonfils} 
\affiliation{CNRS, IPAG, Universit\'e Grenoble Alpes, 38000 Grenoble, France}

\author[0000-0002-9148-034X]{Vincent Bourrier}
\affiliation{Observatoire Astronomique de l'Universit\'e de Gen\`eve, Chemin Pegasi 51b, CH-1290 Versoix, Switzerland}

\author[0000-0001-5383-9393]{Ryan Cloutier}
\altaffiliation{Banting Fellow}
\affiliation{Center for Astrophysics $\vert$ Harvard \& Smithsonian, 60 Garden Street, Cambridge, MA, 02138, USA}

\author[0000-0001-8056-9202]{Jo\~ao Gomes da Silva}
\affiliation{Instituto de Astrof\'{i}sica e Ci\^{e}ncias do Espa\c{c}o, Universidade do Porto, CAUP, Rua das Estrelas, 4150-762 Porto, Portugal}

\author[0000-0001-9796-2158]{Emily Deibert}
\affiliation{David A. Dunlap Department of Astronomy \& Astrophysics, University of Toronto, 50 St. George Street, ON M5S 3H4, Canada}

\author[0000-0001-5099-7978]{Xavier Delfosse} 
\affiliation{CNRS, IPAG, Universit\'e Grenoble Alpes, 38000 Grenoble, France}

\author[0000-0001-5541-2887]{Jean-François Donati}
\affiliation{Universit\'e de Toulouse, CNRS, IRAP, 14 av. Belin, 31400 Toulouse, France}

\author[0000-0001-9704-5405]{David Ehrenreich}
\affiliation{Observatoire Astronomique de l'Universit\'e de Gen\`eve, Chemin Pegasi 51b, CH-1290 Versoix, Switzerland}

\author[0000-0001-8504-283X]{Pedro Figueira}
\affiliation{European Southern Observatory, Alonso de Cordova 3107, Vitacura, Santiago, Chile}
\affiliation{Instituto de Astrof\'{i}sica e Ci\^{e}ncias do Espa\c{c}o, Universidade do Porto, CAUP, Rua das Estrelas, 4150-762 Porto, Portugal}

\author[0000-0003-0536-4607]{Thierry Forveille}
\affiliation{CNRS, IPAG, Universit\'e Grenoble Alpes, 38000 Grenoble, France}

\author[0000-0002-1436-7351]{Pascal Fouqu\'e}
\affiliation{Universit\'e de Toulouse, CNRS, IRAP, 14 av. Belin, 31400 Toulouse, France}
\affiliation{CFHT Corporation; 65-1238 Mamalahoa Hwy; Kamuela, Hawaii 96743; USA}

\author[0000-0002-2592-9612]{Jonathan Gagn\'e}
\affiliation{Plan\'etarium Rio Tinto Alcan, Espace pour la Vie, 4801 av. Pierre de Coubertin, Montr\'eal, QC, Canada.}
\affiliation{Institut de Recherche sur les Exoplan\`etes, Universit\'e de Montr\'eal, D\'epartement de Physique, C.P. 6128 Succ. Centre-ville, Montr\'eal, QC H3C 3J7, Canada.}

\author[0000-0002-5258-6846]{Eric Gaidos}
\affiliation{Department of Earth Sciences, University of Hawai’i at Manoa, Honolulu, HI 96822 USA}

\author[0000-0001-5450-7067]{Guillaume H\'ebrard}
\affiliation{Sorbonne Universit\'e, CNRS, UMR 7095, Institut d’Astrophysique de Paris, 98 bis bd Arago, 75014 Paris, France}

\author[0000-0001-5349-6853]{Ray Jayawardhana}
\affiliation{Department of Astronomy, Cornell University, Ithaca, New York 14853, USA}

\author[0000-0003-0637-5236]{Baptiste Klein}
\affiliation{Sub-department of Astrophysics, Department of Physics, University of Oxford, Oxford OX1 3RH, UK}

\author[0000-0001-7120-5837]{Christophe Lovis}
\affiliation{Observatoire Astronomique de l'Universit\'e de Gen\`eve, Chemin Pegasi 51b, CH-1290 Versoix, Switzerland}

\author[0000-0002-1532-9082]{Jorge H. C. Martins}
\affiliation{Instituto de Astrof\'{i}sica e Ci\^{e}ncias do Espa\c{c}o, Universidade do Porto, CAUP, Rua das Estrelas, 4150-762 Porto, Portugal}

\author[0000-0002-5084-168X]{Eder Martioli}
\affiliation{Sorbonne Universit\'e, CNRS, UMR 7095, Institut d’Astrophysique de Paris, 98 bis bd Arago, 75014 Paris, France}
\affiliation{Laborat\'{o}rio Nacional de Astrof\'{i}sica, Rua Estados Unidos 154, 37504-364, Itajub\'{a} - MG, Brazil}

\author[0000-0002-2842-3924]{Claire Moutou}
\affiliation{Universit\'e de Toulouse, CNRS, IRAP, 14 av. Belin, 31400 Toulouse, France}

\author[0000-0003-4422-2919]{Nuno C. Santos}
\affiliation{Instituto de Astrof\'{i}sica e Ci\^{e}ncias do Espa\c{c}o, Universidade do Porto, CAUP, Rua das Estrelas, 4150-762 Porto, Portugal}
\affiliation{Departamento de F\'isica e Astronomia, Faculdade de Ci\^encias, Universidade do Porto, Rua do Campo Alegre, 4169-007 Porto, Portugal}

\begin{abstract}

We present the first exoplanet atmosphere detection made as part of the SPIRou Legacy Survey, a Large Observing Program of 300 nights exploiting the capabilities of SPIRou, the new near-infrared high-resolution ($R \sim 70\,000$) spectro-polarimeter installed on the Canada-France-Hawaii Telescope (CFHT; 3.6-m). 
We observed two transits of \hdun, an extensively studied hot Jupiter that is known to show prominent water vapor absorption in its transmission spectrum. 
When combining the two transits, we successfully detect the planet's water vapor absorption at 5.9$\sigma$ using a cross-correlation \textit{t}-test, or with a $\Delta$BIC$>10$ using a log-likelihood calculation. 
Using a Bayesian retrieval framework assuming a parametrized T-P profile atmosphere models, we constrain the planet atmosphere parameters, in the region probed by our transmission spectrum, to the following values: \vmr $= -4.4^{+0.4}_{-0.4}$, and \cloud $\gtrsim 0.2$\,bar (grey clouds), both of which are consistent with previous studies of this planet.
Our retrieved water volume mixing ratio is slightly sub-solar although, combining it with the previously retrieved super-solar CO abundances from other studies would imply super-solar C/O ratio. 
We furthermore measure a net blue shift of the planet signal of $-4.62^{+0.46}_{-0.44}$\,\kms, which is somewhat larger than many previous measurements and unlikely to result solely from winds in the planet’s atmosphere, although it could possibly be explained by a transit signal dominated by the trailing limb of the planet. This large blue shift is observed in all the different detection/retrieval methods that were performed and in each of the two transits independently.  
\end{abstract} 


\section{Introduction}
\label{sec:Intro}

The characterization of exoplanet atmospheres using transmission or emission spectroscopy has grown considerably since it was proposed by \cite{Seager2000}. The spectra that such techniques yield can be used to probe the state and composition of an exoplanet's atmosphere. This provides insight into physical and chemical processes at play, which can then be interpreted through different formation pathways and evolutionary histories of the planet, usually by measuring the metallicity and the C/O ratio via their molecular abundances (e.g., \cite{Oberg2011, Pelletier2021}). The most successful observations to date have typically been obtained from space (e.g., \citealt{Madhusudhan2014,Sing2016, Barstow2017,Pinhas2019, Welbanks2019}), with the frequent use of the \textit{Hubble Space Telescope} Wide Field Camera 3 (HST/WFC3) spectrometric and \textit{Spitzer Space Telescope} photometric capabilities. The exquisite image and instrument stability enable the detection of subtle spectral differences induced by the planet's atmosphere during its transit or eclipse and phase curve (which can be on the order of only a few tens of parts-per-million, ppm, e.g., \citealt{Kreidberg2014_WASP43}). One caveat, however, is that these instruments usually have a limited wavelength coverage, which limits the simultaneous detection of water and carbon-bearing molecules (which prevents the computation of an accurate C/O ratio) and can sometimes lead to mixed results, e.g. the unclear nature of the strong spectral feature in the 4.5\,$\mu$m Spitzer band in WASP-127\,b atmosphere, which could come from CO and/or \diox\ \citep{Spake2021}. 
Ground-based high dispersion spectroscopy (HDS) can achieve similar precision by resolving each line independently.
The change in orbital radial velocity of the planet can be exploited to disentangle the signal of the planet from that of stellar and telluric signals (e.g. \citealt{Snellen2010}). In that case, the transit/emission spectrum is probed by looking for a signal (from either atomic or molecular species) that follows the planet radial velocity, which can be orders of magnitudes smaller than that of its star. Compared to low-resolution spectroscopy, HDS has the added benefit of being able to disentangle absorption by different molecules with overlapping bands (but non-overlapping lines, e.g. \citealt{Giacobbe2021}) and from different origins (e.g., stellar, planetary, and telluric) because the individual lines are resolved. Moreover, wind speeds and global dynamics of the planet's atmosphere can be measured (through the Doppler shift and broadening of the planet's absorption lines; e.g., \citealt{Wyttenbach2015, Louden2015, Flowers2019}).

The firsts successful atmospheric characterizations from the ground were detections of sodium in the optical on \hdun\ (\citealt{Redfield2008}; with the High Resolution Spectrograph, $R \sim 60\,000$, on the Hobby-Eberly Telescope) and \hddeux\ (\citealt{Snellen2008}; with the High Dispersion Spectrograph, $R \sim 45\,000$, on the Subaru Telescope; even though these results are inconsistent with the more recent ones from \citet{Casasayas-Barris2020_hd209}).
The first near-infrared (NIR) HDS study was presented by \cite{Snellen2010}, with the detection of carbon monoxide (CO) in the atmosphere of \hddeux\ via transmission spectroscopy using CRIRES ($R \sim 100\,000$; \citealt{CRIRES}) installed at the VLT (8.2-m telescope). Many more such detections followed: \water\ and CO for several transiting and non-transiting exoplanets \citep{Brogi2012, Birkby2013, Brogi2016, Birkby2017, Webb2020}, HCN for \hddeux~\citep{Hawker2018}, and the effects of global atmospheric dynamics, such as day-to-night winds and/or eastward winds (jet streams) on \hddeux\ and \hdun\ \citep{Snellen2010, Brogi2016, Flowers2019, Beltz2021}. 

Multiple other high-resolution instruments have enabled additional interesting results again for transiting and non-transiting targets. A lot of them came from optical instruments (a non-exhaustive list includes HARPS, e.g., sodium detection on WASP-76\,b, \citealt{Seidel2019}; HARPS-North, e.g., neutral Fe and Ti on the atmosphere of KELT-9\,b, \citealt{Hoeijmakers2018}, ESPRESSO, e.g., the iron condensation on the nightside of WASP-76\,b, \citealt{Ehrenreich2020}, and CARMENES VIS-channel, e.g., extended H$\alpha$ envelope detected on KELT-9\,b \citealt{Yan2018}), but the following list highlights some of the near-infrared ones. The NIRSPEC spectrograph (0.95--5.4\,$\mu$m, $R \simeq 25\,000$; \citealt{NIRSPEC}) on the Keck II telescope led to the detection of several species, such as CO, \water\ \citep[e.g., ][]{Rodler2013,Lockwood2014, Piskorz2016,Piskorz2017A,Piskorz2018,Buzard2020}, and more recently metastable helium (He, at 1083\,nm; e.g., \citealt{Kirk2020}). 
CARMENES (NIR-channel, 0.96--1.71\,$\mu$m, $R = 80\,400$, Calar Alto Observatory, 3.5-m; \citealt{CARMENES}) has also been very prolific in recent years. He absorption was detected on multiple targets \citep{Allart2018, Allart2019, Salz2018, Alonso2019_he, Palle2020}, as well as \water\ absorption in both \hdun\ and \hddeux\ \citep{Alonso2019_water, Sanchez-Lopez2019, Sanchez-Lopez2020}. 
Also, GIANO (0.95--2.45\,$\mu$m, $R = 50\,000$, Telescopio Nazionale Galileo, 3.6-m; \citealt{GIANO}) was able to confirm the \hdun\ detection of \water\ \citep{Brogi2018} and of metastable He \citep{Guilluy2020}. Moreover, \cite{Guilluy2019} also detected \water\ absorption in the atmosphere of the non-transiting HD\,102195\,b planet, and found evidence for the presence of methane (\methane) as well. 

The {\em Spectro-Polarim\`etre InfraRouge} (SPIRou; \citealt{Donati2020}) is a new fiber-fed \'echelle spectro-polarimeter operating in the NIR domain installed on the CFHT. SPIRou was primarily designed to detect and characterize Earth-like planets in the habitable zone of low-mass stars via precise radial velocity (RV), down to $\sim$1\,m~s$^{-1}$, and to study the stellar magnetic fields using its polarimetry capabilities \citep[e.g.,][]{Martioli2020, Klein2021}. SPIRou saw its first light April 2018, and is the first instrument to have both a high spectral resolution ($R \sim 70\,000$) as well as such a broad continuous NIR spectral range ($\sim$0.95--2.50~$\mu$m), which provides an unparalleled richness of spectral information. Its wider spectral range gives access to more individual absorption features, thus making SPIRou well suited for exoplanet atmospheric characterization via the cross-correlation technique (see \citealt{Deibert2021}, Pelletier et al., 2021, submitted). SPIRou also has a higher throughput than many slit spectrographs and its design provides a more stable line spread function.

In this paper we report the first results of the SPIRou Legacy Survey (SLS) on exoplanet atmosphere characterization for the planet HD~189733\,b \citep{Bouchy2005}. Initiated at the start of SPIRou operations, the SLS is a CFHT Large Program of 300 telescope nights (PI: Jean-François Donati) whose main goals are to search for planets around M dwarfs using precision RV measurement, characterize the magnetic fields of young low-mass stars and their impact on star and planet formation, and probe the atmosphere of exoplanets using HDS. The planet targeted here, \hdun, is one of the most studied hot Jupiters to date, and due to the brightness of its active host star (K2V; $H = 5.59$\,mag), it offers great opportunities for in-depth analyses. The study of the orbital motion, Rossiter-McLaughlin effect \citep[RME;][]{Rossiter1924,McLaughlin1924} and magnetic field made with these SPIRou SLS observations of \hdun\ are presented in  \cite{Moutou2020}.

Its atmosphere has been studied many times and its transmission spectrum has revealed water \citep[e.g.,][]{Brogi2016, Brogi2018, Alonso2019_water}, CO \citep[][and references therein]{deKok2013_SVD}, hydrogen \citep{Lecavelier2012, Bourrier2013, Bourrier2020}, metastable helium \citep{Salz2018, Guilluy2020}, and sodium \citep{Redfield2008, Jensen2011, Huitson2012, Wyttenbach2015}. The rotation of the planet, consistent with being tidally locked, and evidence of winds have also been detected \citep[e.g.,][]{Wyttenbach2015, Louden2015, Brogi2016, Flowers2019}. Furthermore, \cite{Barstow2020} studied the properties and location of clouds in its atmosphere, which have a substantial effect on retrieved abundances. Their results pointed toward an heterogeneous atmosphere, with small-particle aerosols covering at least 60\% of the terminator region, reaching to low pressures, but no grey cloud.


This paper is organized as follows. In Section~\ref{sec:Obs}, we present the observational setup and describe the observational data along with the details of the reduction. In Section~\ref{sec:analysis} we explain the telluric and stellar signal removal process, and also present the atmospheric models that are used for the cross-correlation analysis. Section~\ref{sec:correl} details the methods that we used to extract the planet's atmosphere signal and present the associated results. In Section~\ref{sec:Discuss} we discuss our findings, and summarize our main results in Section~\ref{sec:conclusion}.

\begin{deluxetable}{lcc}
\tablecaption{SPIRou observations of HD\,189733 \label{tab:param_obs}}
\tablehead{
\colhead{ } & \colhead{Transit 1} & \colhead{Transit 2}
}
\startdata
UT Date                         & 2018-09-22   & 2019-06-15 \\
BJD (d)$^{\,\textrm{a}}$        & 2458383.77   & 2458649.95 \\  
Texp (s)$^{\,\textrm{b}}$            & 250    & 250  \\
Seeing ('')$^{\,\textrm{c}}$         &  0.79     &  0.85    \\
SNR$^{\,\textrm{d}}$                &   259    &  224 \\
\textbf{Number of exposures :} & & \\
Before ingress          &   0   &  12  \\
During transit          &  21   &  24  \\%
After egress            &  15   &  14  \\
Total                   &  36   &  50 \\
Tot. observ. time (h)$^{\,\textrm{e}}$      &  2.50   &  3.47
\enddata
\tablecomments{ \footnotesize
$^{\textrm{a}}$ Barycentric Julian date at the start of the transit sequence
; $^{\textrm{b}}$ Exposure time of a single exposure; $^{\textrm{c}}$ Mean value of the seeing during the transit; $^{\textrm{d}}$ Mean SNR per pixel, per exposure, at 1.7\,$\mu$m; $^{\,\textrm{e}}$ Total observing time in hours.
}
\end{deluxetable}


\section{Observations and data reduction}
\label{sec:Obs}

All data presented here were obtained with SPIRou \citep{Donati2020}. 
The spectrograph covers the $Y$, $J$, $H$, and $K_s$ bands ($\sim$0.95--2.50\,$\mu$m) simultaneously at a nominal resolving power of $R = \lambda/\Delta\lambda \sim 70\,000$ (with a sampling precision of $\sim 2.3$\,\kms\ per pixel), over 50 spectral orders. 
The fiber-fed spectrograph is bench-mounted in a vacuum tank to maximize its stability and radial velocity precision. This further gives a much more stable line-spread function, greatly reducing the level of systematic errors \citep{Artigau2014}.
SPIRou has a $4096 \times 4096$-pixel H4RG detector, with $15\,\mu$m-wide pixels.
The overall throughput is around 4 to 8\% in the $Y$ and $J$ bands, while it increases to 10--12\% in the $H$ and $K$ bands. There are two science fibers that monitor the object and one calibration fiber that can track a Fabry-P\'{e}rot etalon for simultaneous wavelength calibration. 

\begin{figure}
\includegraphics[width=\linewidth]{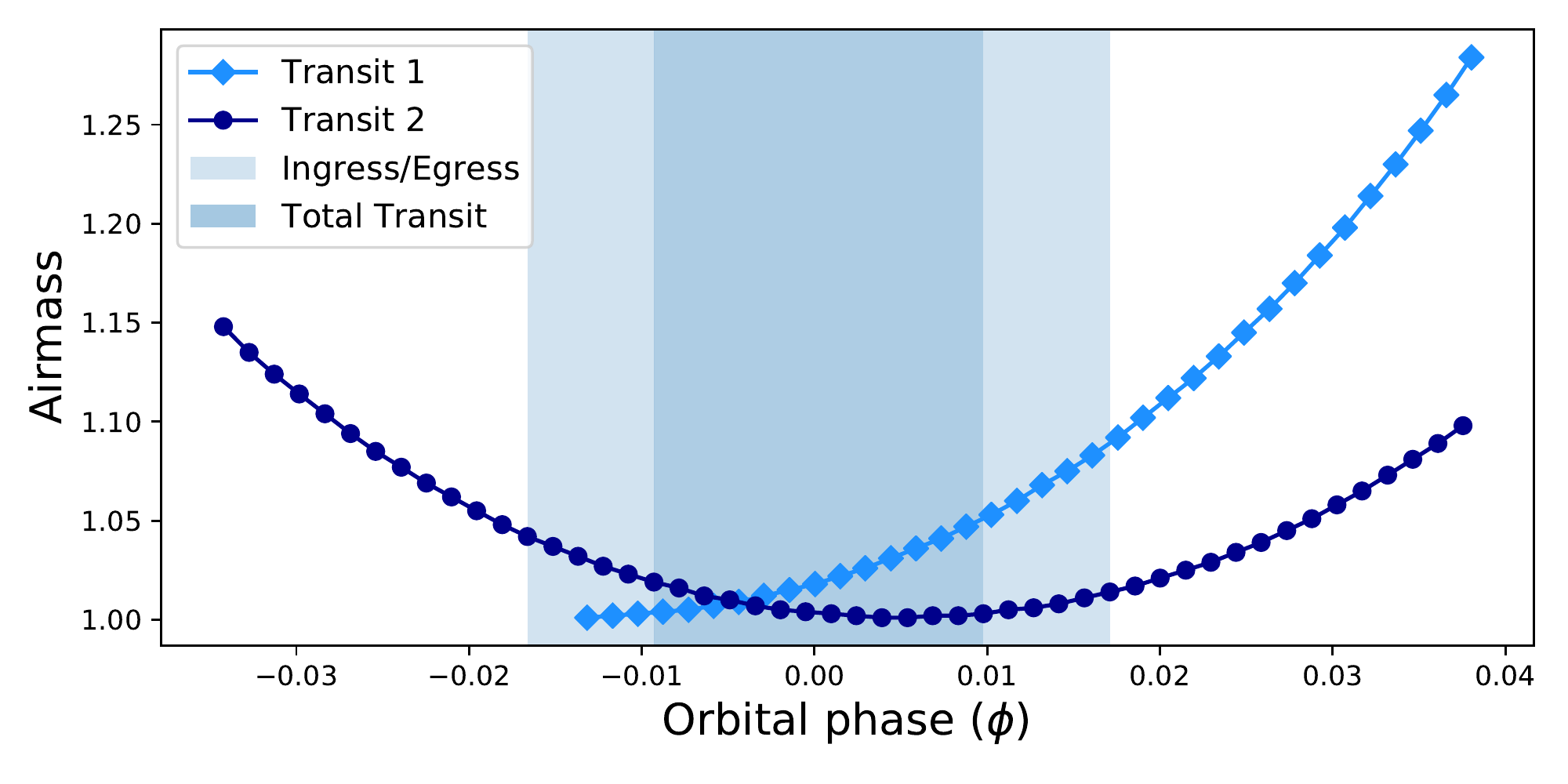}
\caption{\label{fig:AM}
Airmass variation during the two SPIRou transit observations (light blue diamonds for Tr1 and dark blue circles for Tr2; the shaded area shows the span of the transit event).
} 
\end{figure}

\begin{figure}
\includegraphics[width=\linewidth]{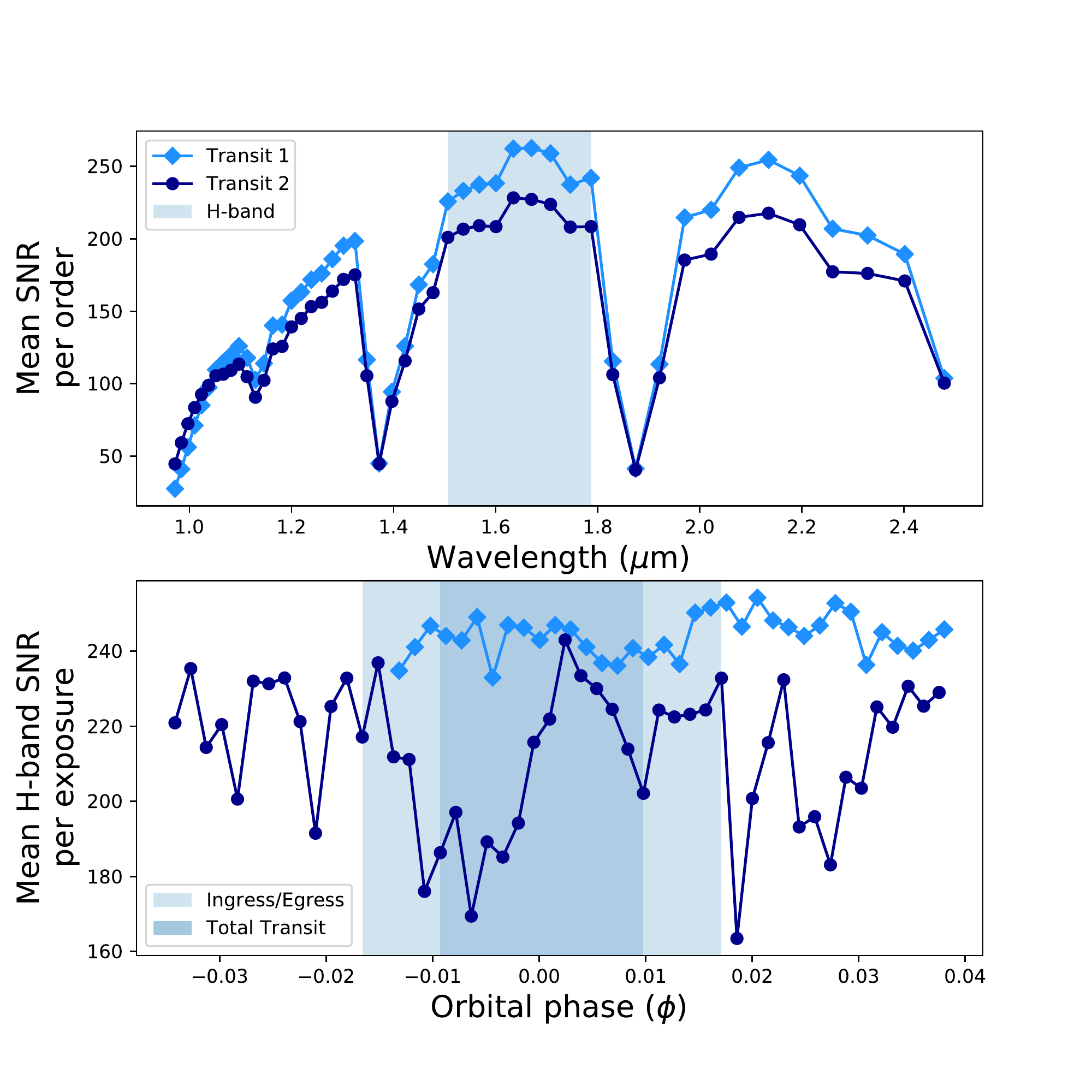}
\caption{\label{fig:SNR}
Temporal mean of the SNR per order (top panel; the shaded area highlights the orders in the $H$-band) and spectral mean of the SNR for $H$-band per exposure (bottom panel; the shaded area shows the span of the transit event) during SPIRou observations (light blue diamonds for Tr1 and dark blue circles for Tr2).
} 
\end{figure}

Two transits of \hdun\ were observed as part of the SLS. The first transit (hereafter Tr1) was observed on UT September 22, 2018, as part of SPIRou commissioning observations (later combined to SLS data), and the second (hereafter Tr2) on UT June 15, 2019. Both sets of observations were taken without moving the polarimeter optics (rhombs) to ensure the highest possible instrument stability, with the Fabry-P\'{e}rot in the calibration channel, and both with an exposure time of 250\,s per individual exposure. Table~\ref{tab:param_obs} lists the parameters of the observations. 
The first data set consists of 2.5\,h, divided into 36 exposures, where the first 21 are in-transit and the remaining 15 are out-of-transit. Technical operations during this night prevented the start of the sequence early enough to observe the star before ingress. The second data set consists of 50 exposures in total, where 24 are in-transit, 12 before, and 14 after transit, for a total of $\sim 3.5$\,h.
Conditions were photometric for both transit sequences, with an average seeing of around 0.82\arcsec\ as estimated from the guiding images. The airmass remained under 1.3 for the total duration of both observations (see Figure~\ref{fig:AM}). 
The water column density was more stable during the second night with a range of 2.5--3.0 vs 2.6--3.6\,mm \water\ for the first night. 
The signal-to-noise ratio (SNR) temporal mean per order and the spectral mean (over the $H$-band only) per exposure for both transits are shown in Figure~\ref{fig:SNR}. The Tr2 SNR is much more variable, but has out-of-transit observations on both sides.

The data were reduced using \texttt{APERO} (A PipelinE to Reduce Observations; version 0.6.131; Cook et al., in prep.), the SPIRou data reduction software.

The first step is to pre-process the raw data. This removes certain detector effects (correcting for the top and bottom reference pixels, median filtering against the dark amplifiers and the \texttt{1/f} noise; \citealt{Artigau2018}). \texttt{APERO} then calibrates observations, correcting for the dark, flagging bad pixels (from a bad pixel map created from calibration flats and darks), removing background scattered light (both locally and globally), and cleaning hot pixels (via interpolating over high-sigma outliers compared to their immediate neighbors). 

The order positions are found and fit with polynomials and pixels are registered onto a common reference grid (using a master Fabry-P\'{e}rot, FP). In addition, the order geometry and the geometry of the detector (slicer shape, slicer tilt and other optics) are separated into changes across-order, along-order and an affine transformation matrix (essentially being characterized by a shift in dx, dy and an A-B-C-D matrix, which can describe a translation, reflection, scale, rotation and or shear in the detector). These geometric changes are applied, along with the order polynomial fits in order to straighten the image.  

The straightened image is then optimally extracted \citep{Horne1986} and cosmic ray correction is applied. This produces an extracted 2D spectrum; $E2DS$ of dimensions 4088 (4096 minus 8 reference pixels) by 49 orders (corresponding to physical orders of \#79 in the blue to \#31 in the red). The $E2DS$ are flat and blaze corrected (the blaze is fitted with a simple $sinc$ model) and a wavelength solution is available for each night (the wave calibration is done using both a hollow-cathode UNe lamp and the FP; \citealt{Hobson2021}). Thermal background is corrected with a dark calibration, scaled in amplitude to match the thermal background of the reddest orders. 

As a final step, when the FP is used simultaneously in the calibration fiber, the contamination from the simultaneous FP is corrected for (using a set of calibrations with dark in the science fiber and the FP in the reference fiber to correct the contamination from the reference fiber into the science fiber). While the $E2DS$ are produced for the two science fibers (A and B) and for the combined flux in the science fibers (AB), we only used the AB extraction as this is the relevant data product for non-polarimetric observations.

\section{Analysis}
\label{sec:analysis}

\begin{figure*}[tbph]
\centering
\vspace*{-2cm} 
\includegraphics[width=\linewidth]{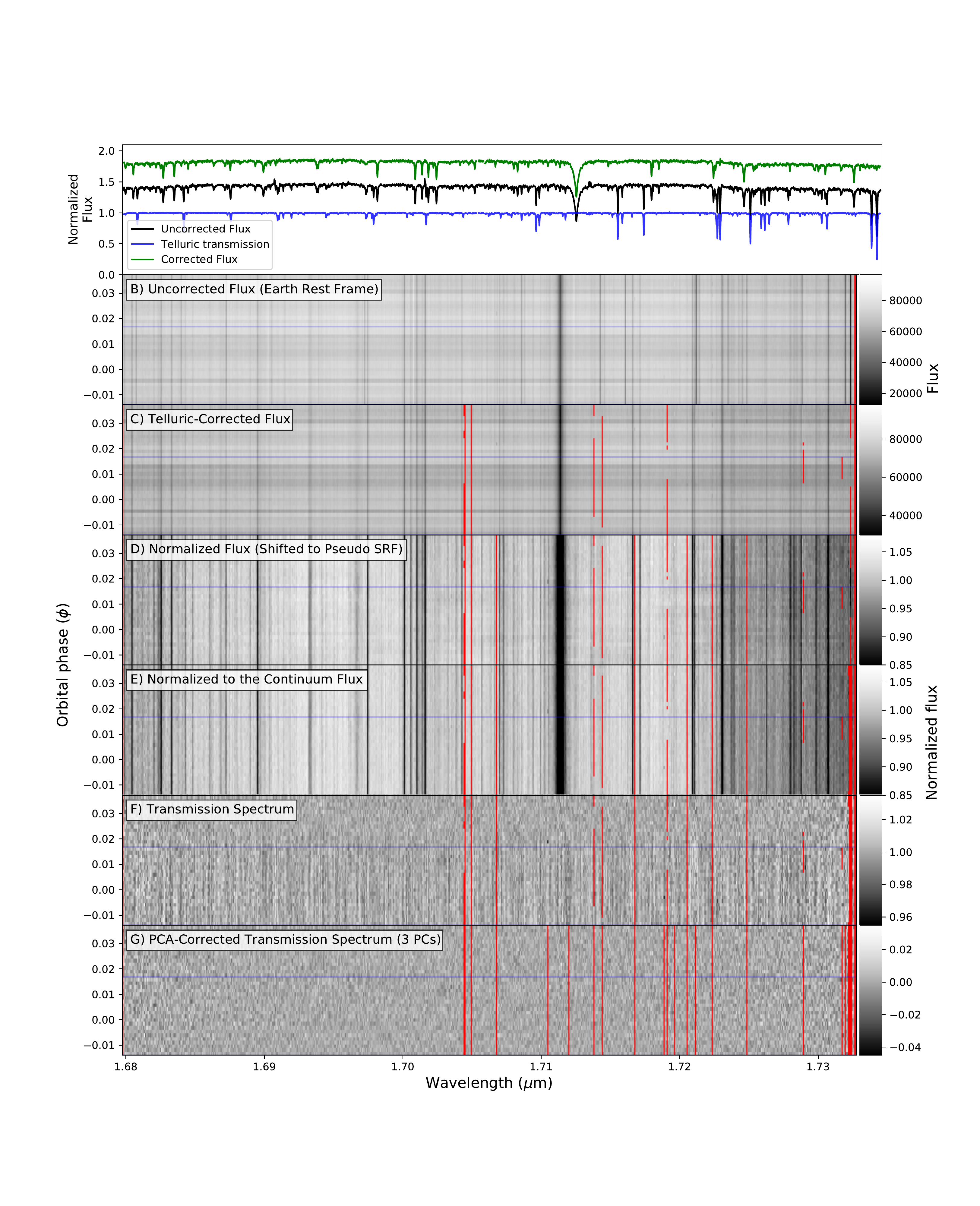}
\vspace{-50pt} 
\caption{\label{fig:steps}
Analysis steps that are applied to the observed spectral time series of the first transit. Here the full order spanning 1.6797 to 1.7327~$\mu$m is shown. 
\textit{Panel A}: The uncorrected (black), the telluric-corrected (green) observed spectra and the reconstructed telluric transmission spectrum (blue) are shown with an offset to facilitate visibility. 
\textit{Panel B}: Uncorrected spectra (counts normalized by the blaze function).
\textit{Panel C}: Telluric-corrected spectra. The masked pixels are shown in red; here the masking is done by \texttt{APERO}.
\textit{Panel D}: The high variance columns of the mean normalized spectra are masked, and then the spectra are co-aligned (shifted) in the pseudo-stellar rest frame.
\textit{Panel E}: The regions with deep telluric lines are masked. Every spectrum is normalized to the continuum level of the Master-Out spectrum.
\textit{Panel F}: Planetary transmission spectra, where each spectrum was divided by the Master-Out spectrum.
\textit{Panel G}: Final planetary transmission spectra corrected for the vertical residual structures using PCA. 
The blue line in panels B-G shows the egress position (mid-transit is at phase 0).}
\end{figure*}

\subsection{Telluric absorption correction}
\label{subsec:tell_rmv}

\begin{deluxetable*}{lcccr}
\tablecaption{Adopted parameters for the system HD\,189733 \label{tab:param_sys}}
\tablehead{
\colhead{Parameter} & \colhead{Symbol} & \colhead{Value} & \colhead{Unit} & \colhead{Reference}
}
\startdata
Stellar mass & $M_{\star}$ & $0.806 \pm 0.048$ & $M_{\odot}$ & T08, B19 \\ 
Stellar radius & $R_{\star}$ & $0.756 \pm 0.018$ & $R_{\odot}$ & T08 \\
Planet mass & $M_{\rm p}$ & $1.123 \pm 0.045$ & $M_{\rm J}$ & T08 \\ 
Planet radius & $R_{\rm p}$ &  $1.138 \pm 0.027$ & $R_{\rm J}$ & T08 \\ 
Semi-amplitude & $K$ &  $0.205 \pm 0.006$ & \kms & T08 \\ 
Orbital semi-major axis & $a$ & $0.03099^{+0.00060}_{-0.00063}$ & AU & T08 \\
Orbital period & $P_{\rm orb}$ & $2.218575123 \pm 0.000000057$ & d & B19 \\ 
Orbital eccentricity & $e$ & $0.0028 \pm 0.0038$ &  & B19 \\ 
Orbital inclination & $i_{\rm P}$ & $85.712 \pm 0.036$ & $^{\circ}$ (deg) & B19 \\
Epoch of transit & $T_0$   &  $2453968.837031 \pm 0.000020$ & BJD &  B19 \\ 
Transit duration & $T_{14}$ & $1.8012 \pm 0.0016$ & h & B19 \\ 
Systemic velocity Tr1 & $v_{\rm sys, Tr1}$ & $-2.59 \pm 0.21$ & \kms & This work \\ 
Systemic velocity Tr2 &$v_{\rm sys, Tr2}$ & $-2.76 \pm 0.21$ & \kms & This work 
\enddata
\tablecomments{(T08) \citealt{Torres2008} and (B19) \citealt{Baluev2019}.\\  
} 
\end{deluxetable*}


A large portion of the NIR domain that SPIRou covers is affected by absorption in the Earth's atmosphere (telluric absorption), which is resolved into thousands of individual narrow molecular lines at SPIRou's resolution. This absorption is strongest and line densities are highest in between the $YJHK$ photometric bandpasses, but many lines are nevertheless present with various strengths throughout the domain. It is crucial that these telluric absorption lines be precisely corrected for or masked out prior to seeking the subtle spectral signature of an exoplanet atmosphere seen in transit, especially as they arise from molecules also expected to be present in exoplanet atmospheres.

The spectra are corrected for telluric line contamination within \texttt{APERO} in a two-step process.
First, the blaze normalized $E2DS$ spectra are pre-cleaned by removing the best-fitting TAPAS atmospheric transmission spectrum \citep{TAPAS}. 
This best-fit model is found by minimizing the cross-correlation signal between the TAPAS model and the telluric residuals, i.e. the data from which this same model was removed. 
Second, the telluric line residuals are corrected using the Principal Component Analysis (PCA) approach developed by \citet{Artigau2014} and implemented in \texttt{APERO} (Artigau et al., in prep.). The approach exploits the fact that the absorbance spectrum of Earth's atmosphere can be expressed, in log-space, as a linear combination of absorbance spectra from different chemical species (chiefly H$_2$O, O$_2$, CO$_2$ with minor O$_3$, CH$_4$ and N$_2$O features). As part of SPIRou night time calibrations, rapidly rotating A stars are observed and the derived telluric absorption from these observations is added to a library. The telluric correction spectrum is constructed through a linear combination (in log space) of the first 7 principal components (PCs) of the library of telluric absorptions, for each observed spectrum. 

The most opaque regions in Earth's atmosphere (saturated lines, or lines with transmission smaller than 10\%) will let little to no light through, resulting in a poorly constrained/unreliable correction. The \texttt{APERO} pipeline performs the telluric absorption correction for lines with a transmission down to $\sim 10$\%, and deeper lines are masked out. 
The total reconstructed Earth transmission spectrum calculated by the pipeline (the product of the best-fit TAPAS model and the PCA reconstruction of the residuals) is one of the data product, allowing further masking if desired. 
An example of the telluric absorption reconstruction and correction is shown in Figure~\ref{fig:steps}, panels A to C.

Sky emission lines are also removed by the \texttt{APERO} pipeline as part of the telluric correction process. The pipeline draws from a library of sky spectra taken at various times through the life of SPIRou and constructs a linear combination of the first 9 PCs of this library, again for every exposure individually, to reconstruct and subtract the sky emission lines in the observed spectrum. The final telluric corrected spectra are re-multiplied by the blaze to keep the level proportional to the photon counts.

In our analysis below, we found that further masking of deep telluric lines (i.e. further masking parts of the previously telluric corrected regions) improved the results, possibly because low-level telluric absorption residuals remain in the spectra. Using the reconstructed telluric spectrum to define our mask, we masked all telluric lines for which the transmission in the core is below 30\% in Tr1 and 35\% in Tr2, and for those lines, we also mask the wings by extending the mask until the transmission reaches 97\% (Figure~\ref{fig:steps}\,E); these limits were empirically determined, based on an injection-recovery test. This step removes a total 8.6\% and 10.6\% of the spectral domain in Tr1 and Tr2, respectively. This masking is applied at step 3 in the following section.

\subsection{Transmission spectrum construction}
\label{subsec:transpec}

Starting from the telluric-corrected $E2DS$ spectra, after division by the normalized blaze function to flatten them, we applied the following operations to construct the planet's transmission spectra. These operations were applied on each order individually and separately for each transit sequence.

1) To remove bad pixels, the spectra are first normalized by dividing them by their median value (over wavelengths). Then, spectral pixels with a time variance (over the different spectra) more than $4\sigma$ above the mean variance are masked. 
In total, this removes 0.6\% and 0.5\% of the pixels in Tr1 and Tr2, respectively. 

2) All the spectra are Doppler shifted (via cubic spline interpolation that handles masked arrays) from the observer to a pseudo-stellar rest frame (SRF) by the opposite of the stellar radial velocity variation relative to the middle of the sequence, $\Delta v_{\rm S}(t)$. The stellar radial velocity $v_{\rm S}(t)$ is itself defined as,

\begin{equation}
v_{\rm S}(t) = v_{\rm bary}(t) + v_{\rm sys} + v_{\rm reflex}(t) \,,
\end{equation}

where $v_{\rm bary}$ is the barycentric velocity of the observer (and in our case it is the barycentric Earth radial velocity, BERV), \vsys\ is the systemic radial velocity, and $v_{\rm reflex}(t) = K \sin [2\pi(\phi(t)+0.5)]$ is the radial component of the reflex motion of the star induced by the planet (Keplerian), where $K$ is the stellar radial velocity semi-amplitude (which we fixed; see Table~\ref{tab:param_sys}), and $\phi$ is the planet orbital phase ($\phi = 0$ at mid-transit). We then define $\Delta v_{\rm S}(t)$ as follows:

\begin{equation}
\Delta v_{\rm S}(t) = \Delta v_{\rm bary}(t) + \Delta v_{\rm reflex}(t) \,,
\end{equation}

where $\Delta v_{\rm bary}(t) = v_{\rm bary}(t)- v_{\rm bary}(t_{\rm mid. exp.})$, the difference between the barycentric velocity during the sequence ($v_{\rm bary}(t)$) and its value at the middle of the sequence ($v_{\rm bary}(t_{\rm mid. exp.})$), and equivalently for the reflex motion term. 
This aligns the stellar lines across all spectra (see Figure~\ref{fig:steps}\,D), even though they are not exactly in the SRF, while minimizing the shifts applied to the data, hence minimizing the associated interpolation errors (the first half of the exposures are thus shifted by roughly the same amount as the second half, but in the opposite direction). 
In our case, the shifts applied are at most $\sim 5\%$ and $\sim 7\%$ of a SPIRou spectral pixel, for Tr1 and Tr2 respectively. 
Many of the HD~189733 system and orbital parameters are well known and the values that were used here are listed in Table~\ref{tab:param_sys}. We measured \vsys\ directly from our data by computing the CCF of the telluric corrected spectra with a synthetic spectrum from a PHOENIX atmosphere model \citep{Husser2013} with $T_{\rm eff} = 5100\,$K, $\log g = 4.5$, and $Z=0$. We computed this CCF for all orders across the $H$ band, measured the peak position, subtracted $v_{\rm bary}(t)$ and $ v_{\rm reflex}(t)$, then calculated the mean value by weighting by the SNR of the orders, and finally computed the mean over all spectra for each transit. We obtained values of $-2.22 \pm 0.04$\,\kms\ for Tr1, and $-2.39 \pm 0.04$\,\kms\ for Tr2. Then to determine the heliocentric radial velocity of the HD~189733 system from our observations, we subtracted $0.67 \pm 0.04$\,\kms\ to compensate for the gravitational redshift and $-0.3 \pm 0.2$\,\kms\ for the convective blueshift of a K2 star \citep{Leao2019}. The final $v_{\rm sys}$ values are reported in Table~\ref{tab:param_sys}. 

3) The additional telluric masking mentioned above is applied; this was not done earlier to limit interpolations on masked pixels (Figure~\ref{fig:steps}\,D to E).

4) The stellar spectral features are removed following the technique of \cite{Allart2017}: (a) A ``Master-Out'' spectrum is built by taking the mean of all the out-of-transit spectra (full-disk stellar spectrum). (b) All spectra are divided by a low-pass-filtered\footnote{Median filter of width 51 pixels followed by convolution with a Gaussian kernel of width 5 pixels.} version of their ratio with the Master-Out; this flattens out the continuum and removes the modal noise, thus reducing the correlation noise due to slopes in the spectra (Figure~\ref{fig:steps}\,E). (c) A second iteration of the Master-Out spectrum is calculated using the normalized out-of-transit spectra. (d) The normalized spectra are divided by this final Master-Out spectrum, which yields the planetary transmission spectra (see Figure~\ref{fig:steps}\,F), and sigma clipped at 6\,$\sigma$, to prevent outliers to be included in the next step.

5) At this stage, we can see residual vertical features in the spectral time series 2D representation (Figure~\ref{fig:steps}\,F); these seem to appear because of the Master-Out is built with a subset of spectra and better represents the spectra closest to it in time (the out-of-transit spectra, in this case), and the residuals could be linked to systematic effects that vary with time. 
To remove these residual features, we use a PCA approach. 
The PCs are built from the logarithm of the flux in the time dimension using each spectral pixel time series as a sample point (which is the transpose equivalent to building the PCs basis in the spectral space and having each exposure as a sample point).
Each sample (spectral pixel time series) is then divided by the exponential of their reconstruction based on the firsts $n$ chosen PCs.  
To determine the number of PCs to use, we performed an injection-recovery test for both analysis methods presented below, the cross-correlation function (CCF) and the log-likelihood.
We injected the model at -\kp\ and the known \vsys\ + \vv{rad}.
This test indicates that removing 2 and 3 PCs yields the best detection significance of the injected signals for Tr1 and Tr2, respectively, and this is what we adopted for all analyses. The higher telluric fraction to be masked and the number of PCs to remove in the second transit could be linked to its lower data quality compared to Tr1.

6) Finally, any remaining outlier pixels are masked using sigma clipping (at $3 \sigma$) in the time dimension, and the mean of each spectrum is subtracted out, to keep a zero mean for the cross-correlation. This yields the final spectral planetary transmission values $f_i$ (where $i$ indexes over both time and wavelength), shown in Figure~\ref{fig:steps}\,G, that are used for the cross-correlation/log-likelihood mapping in Section~\ref{sec:correl}.

During the analysis of the Tr2 data set, it was found that a significant change in systematic spectral noise occurred around passage of the target through the meridian, near airmass of 1. Similar noise patterns, always occurring as the target passes near the zenith, were also found in other SPIRou data sets (e.g. \citealt{Pelletier2021}). The cause of this is still being investigated, but is thought to be related to the azimuth angle of the telescope (TelAz). This effect is most important when the variation (or gradient) of the TelAz is maximal. See \cite{Pelletier2021} for a longer discussion of this effect.
In our case, this effect had no impact on Tr1, as all data were taken after meridian crossing. For Tr2, we managed to work around this problem by splitting the data set in two subsets: before (27 exposures) and after (23 exposures) the meridian crossing. We applied all the analysis steps above to both subsets independently, and then we merged them back into a single sequence at the end.

\subsection{Atmospheric Model}
\label{subsec:model}
Obtaining information about the molecular content in the planet's atmosphere requires the use of the cross-correlation method/log-likelihood mapping with high spectral resolution synthetic planetary transmission spectra. 
The models of \hduns and associated transmission spectra that were used here were generated using the SCARLET framework \citep[Self-Consistent Atmospheric RetrievaL framework for ExoplaneTs; ][]{Benneke2012, Benneke2013, Benneke2015, Benneke2019_Nature}. SCARLET generates transmission spectra for a simulated planetary atmosphere in hydrostatic equilibrium by considering the molecular opacities at each pressure layer. 
Included in the opacities are contributions from H$_2$O, H$_2$ broadening following \citet{Burrows2003}, and collision-induced broadening from H$_2$/H$_2$ as well as H$_2$/He collisions from \citet{Borysow2002}. 
The models are generated at a resolving power of $R=250\,000$ using line-by-line radiative transfer and are later convolved to match the instrumental resolution of SPIRou. The resulting output is the transit depth as a function of wavelength, i.e.\ $R_p^2(\lambda)/R_\star^2$, which corresponds to the observed transmission spectrum calculated above, pending normalization by the continuum level.

The choice of the water lines list used in the models is important, given the preponderance of this molecule in the atmosphere and throughout the transmission spectrum of \hdun. In this work, we adopted the POKAZATEL \citep{POKAZATEL} water line list from Exomol \citep{Tennyson2016}. 
Most previous analyses used the HITEMP 2010 line list, but \cite{Gandhi2020} explicitly compared the two with \hdun\ CRIRES data from \cite{Birkby2013} (thermal emission) and reported a strong agreement between the two, with a slightly higher signal from POKAZATEL. Similarly, \cite{Nugroho2021} observed a better detection significance for \water\ in WASP-33\,b when using the POKAZATEL line list than when using HITEMP 2010, even though the signal is found at the same location. \cite{Webb2020} also compared these line lists in their study of the non-transiting HD~179949\,b's atmosphere and found consistent results, although their data favored HITEMP 2010 models.  

In principle, any temperature-pressure (T-P) profile can be used to generate the atmospheric models, but here we adopted an analytical atmospheric T-P profile from \cite{Guillot2010}. For simplicity, we fixed three of the four parameters of the profile, namely $\kappa_{\rm IR}$ the atmospheric opacity in the IR wavelengths, $\gamma$ the ratio between the optical and IR opacity, and $T_{\rm int}$ the planetary internal temperature, while keeping $T_{\rm eq}$, the atmospheric equilibrium temperature, as a free parameter. We fixed the values to $\kappa_{\rm IR} = 10^{-1.5}$, $\gamma = 10^{-0.85}$ and $T_{\rm int} = 100$ so that the shape of the resulting profile would roughly resemble those found in the literature for HD~189733\,b, more specifically from \cite{Sing2016, Brogi2019}.
Both hazes and a grey cloud deck can be included in the models, but here we neglected the hazes contribution given the NIR wavelength range of SPIRou. We did, however, include a grey cloud deck contribution, characterized by its cloud top pressure \cloud\ (bar), as this can have a large effect on the contrasts/depths of spectral lines. Rayleigh scattering is included by default even though its contribution is not significant in the NIR.

The models considered in our study are thus described by 3 free parameters: the water VMR, the temperature \tp\ (K) of the isothermal atmosphere profile, and the cloud top pressure \cloud\ (bar). 
Once a model is generated, we convolve it to the resolving power of SPIRou, 70\,000, and bin it to match the observed data pixel sampling. 

\section{Atmospheric Signal Extraction:\\ Methods and results} 
\label{sec:correl}

Even with a high SNR SPIRou spectrum, individual planetary absorption lines are often weak and buried in the noise. To maximize the signal and the detection strength, we combine the signal from many lines via the cross-correlation and log-likelihood mapping techniques. This is why a maximum number of spectral lines is desired, which then justifies the need for a wide spectral range. In this work, we tested three specific approaches for detecting HD\,189733\,b's signal and constraining its atmospheric parameters, but the models first need to be processed to better represent the data. In this section, we present how we process the models followed by the three approaches, i.e. the cross-correlation (and \textit{t}-test), the log-likelihood mapping and MCMC retrieval.

\subsection{Model processing}
\label{subsec:mod_seq}

At this point in the analysis, the planetary atmosphere signal was affected by all the processing steps that were applied to the data. 
This mostly concerns the subtraction of the PCs, which are generally not orthogonal to the planet transmission spectra. Their subtraction (step 5 from Section~\ref{subsec:transpec}) may remove part of the actual planet signal and introduce artefacts in the spectral time series, which can then bias the determination of the best parameters (velocities, abundances, temperature, etc.). 
We thus need to apply the same treatment to the model before comparing it to the data to ensure a better representation. So instead of using the models directly for the cross-correlation, we use a processed model transit sequence. We proceed as follows:  
we generate full synthetic transit sequences by injecting the model spectrum (described in Section~\ref{subsec:model}) in a reconstruction of the observed data. This reconstructed signal is built by multiplying together (i) the spectral median of every spectrum (for every order; same as step 1 from Section~\ref{subsec:transpec}), (ii) the master-out spectrum and (iii) the PCA reconstructed version of the transmission spectrum, all of which are (mostly) planet signal free. 
We inject the model at \vv{P}$(t)$, the total planet radial velocity,

\begin{equation}
v_{\rm P}(t) =  v_{\rm bary}(t) + v_{\rm sys} + v_{\rm reflex, P}(t) - v_{\rm reflex}(t) +  v_{\rm rad}\,,
\label{eq:v_planet}
\end{equation}

where $v_{\rm reflex, P}(t) = K_{\rm P} \sin [2\pi(\phi(t))]$ is the radial part of orbital velocity of the planet and \kp\ is the planet's radial velocity semi-amplitude, and $v_{\rm rad}$ is a constant additional velocity term to account for potential shifts. 
Since the spectra are in the pseudo-SRF (from step 2), here \vv{bary}$(t) =$\,\vv{bary}$(t_{\rm mid. exp.})$.
We can inject the models at different combinations of \kp\ and \vv{rad}. 
Then, the relevant steps of the analysis are reapplied on the modeled transit sequence, i.e.\ 1, 5 and 6;  step 2 is ignored since the reconstructed signal is already in the pseudo-SRF, so no additional shifts are needed; step 3 is ignored since the spectra are already masked (from the reconstructed spectra) so no additional masks are applied (this also excludes the sigma clipping parts in steps 1 and 6); and step 4 is ignored since the spectra are already normalized to the Master-Out continuum, and the Master-Out stays the same. 
For step 5, we project the PCs obtained with the real observations onto the synthetic transit sequence, and remove this reconstruction from the synthetic sequence. The effect of this process is to replicate in the model spectra, to the extent possible, the subtraction of the planet signal that occurs when subtracting the PCs from the actual data. After this PCs subtraction, we then remove the mean light curve for each order, as was done on the observed data. This synthetic, PCs- and mean-subtracted time series ($m_i$) is then used to calculate the cross-correlation and log-likelihood. 


\subsection{Cross-Correlation}
\label{subsec:CCF}

The transmission model of the planet's atmosphere can be cross-correlated with the observed planetary transmission spectra. If the model is adequate and the molecules in the model are present in the planet's atmosphere, then there should be a peak in the cross-correlation function (CCF) at the expected radial velocity of the planet, and, in subsequent exposures that peak should shift according to the orbital velocity of the planet. From a sequence of several exposures in transit, we can then isolate the signal from the planet by combining the correlation signal that comes only from the right radial velocity in each exposure.

\subsubsection{Algorithm} 

Based on the equations in \cite{Gibson2020} (and also used in \citealt{Nugroho2021}), the CCF can be written as,

\begin{equation}
\mathrm{CCF}(\boldsymbol{\theta}) = \sum^N_{i=1} \frac{ f_i \cdot m_i(\boldsymbol{\theta})}{\sigma_i^2}\,,  
\label{eq:CCF}
\end{equation}

which is equivalent to a weighted CCF, where $f_i$ are the observed values of the planetary transmission spectrum (described in Section~\ref{subsec:transpec}) with associated uncertainties $\sigma_i$, $m_i$ are the model values (described in Sections~\ref{subsec:model} and \ref{subsec:mod_seq}), and $\boldsymbol{\theta}$ is the model parameter vector, which includes the atmospheric model parameters and the applied orbital and systemic velocities (i.e.\ the Doppler shift at velocity \vv{P}(t)). The index $i$ represents both time and wavelength, and the summation is done over $N$ data points (number of unmasked pixels, $N=4\,564\,944$ for Tr1 and $N=6\,656\,359$ for Tr2). In practice, we first sum over wavelengths for each order, then over all orders, then over time. When summing over time, we apply a weighing to each spectrum according to a transit model of \hdun, which takes into account that the planet's signal is present only during the transit, while also being weaker during the ingress and egress. To build the transit model (transit depth at a given time), we use the equations from \cite{Mandel2002}, assuming a 4-parameters non-linear limb-darkening law \citep{Claret2000} with theoretical coefficients $u_1 = 0.9488$, $u_2 = -0.5850$, $u_3 = 0.3856$, and $u_4 = -0.1318$, based on 3D models of HD~189733 taken in \cite{Hayek2012} and valid for the H-band\footnote{It was computed with the \texttt{BATMAN PYTHON} package \citep{Kreidberg2015_batman}.}. We note that using a variable limb-darkening law for the different parts of the spectrum would be more precise, but refrain from doing so for simplicity. In fact, going from a simple linear law to a non-linear one only had a minimal impact on the results. We used the ephemeris and system parameters listed in Table~\ref{tab:param_sys}.

The uncertainties $\sigma_i$ were determined by first calculating, for each spectral pixel, the standard deviation over time of the $f_i$ values. However, to make sure we do not underestimate the noise by removing too many PCs, we take the standard deviation of $f_i$ after the removal of only one PC, even though the final transmission spectra are corrected from three.
This provides an empirical measure of the relative noise across the spectral pixels that captures not only the variance due to photon noise but also due to such effects as the telluric and background subtraction residuals, but it does not convey how the noise inherent to one spectrum compares to that of another. To capture this latter effect and include it in the $\sigma_i$, the dispersion values calculated above were multiplied, for each spectrum, by the ratio of the median relative photon noise of that spectrum divided by the median relative photon noise of all spectra. This is computed prior to normalization : the SNR variations across the night and the different orders are thus accounted for in this $\sigma_i$ term, acting as a weight. The final uncertainty values $\sigma_i$ thus reflect both temporal and spectral variability.

The CCF (equation \ref{eq:CCF}) is calculated for every order of every spectrum in the transit sequence (time series) for an array of \vv{rad} of size $n_v$. This gives a $86 \times 49 \times n_v$ matrix (when combining the transits\footnote{We combine the two transits by simply concatenating the two CCF time series, order-per-order. There are 36 spectra in Tr1 and 50 in Tr2, for a total of 86.}) for a given \kp\ value and for each model tested. We then take the sum of the CCFs over all of the available SPIRou spectral orders, which gives a 2D cross-correlation map ($86 \times n_v$) as a function of time and \vv{rad}.

\begin{figure}
\hspace*{-0.4cm}
\includegraphics[scale=0.50]{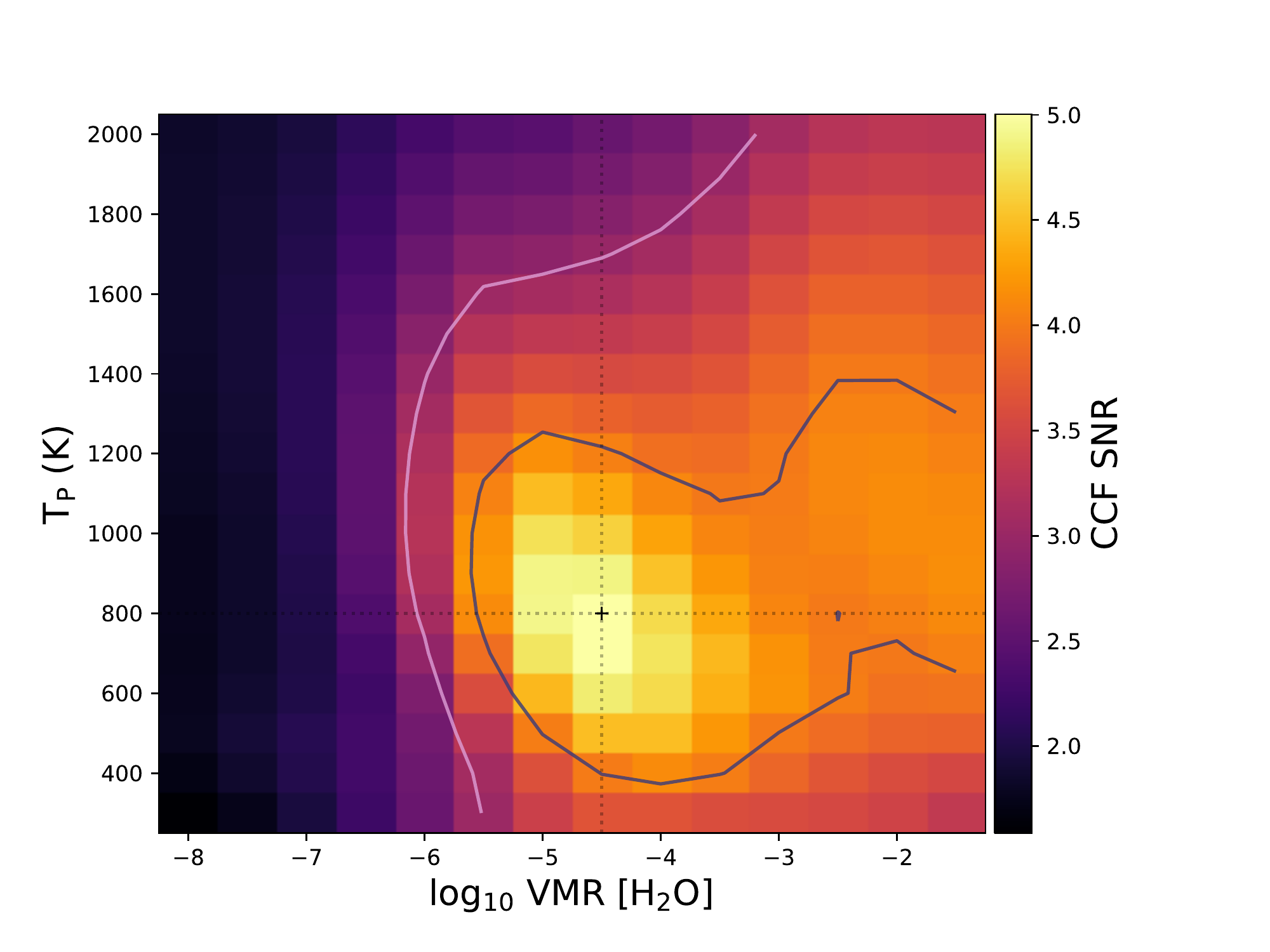}
\caption{Maximum SNR of the CCF of different models as a function of \vmr\ and \tp\, for \cloud\ $ = 10^{-0.5}$\,bar, \kp\ $ = K_{\rm P, 0}$ and \vv{rad} $ = v_{\rm peak}$, using the combined transits. The contour shows where the SNR drops by 1 and 2 from the maximum. A maximum SNR of 5.0 is found at $\log_{10} $VMR[\water]$ = -4.5$, \tp $ = 800$\,K, and \cloud$ = 10^{-0.5}$\,bar.}
\label{fig:grid_ccf}
\end{figure}

We observed that a few spectra, which happened to be those where the absolute value of the TelAz angle gradient was near its maximum value (at meridian passing), had noisier CCFs than the others. We chose to exclude the two spectra most affected by this, corresponding to exposures \#26 and 27 of Tr2 (with TelAz angle gradient above $18^\circ$-per-exposure; Tr1 is not affected by this as the star did not cross the meridian). This exclusion was also applied to the analysis in the following sections.


From the total CCF as a function of the \vv{rad} (again, for a given \kp\ and a given model), we compute the SNR by dividing the total CCF by its standard deviation, the latter being calculated by excluding the region around the peak ($\pm 15$\,\kms).
We expect the correlation peak to follow \vv{rad}$=0$\,\kms\ when $K_{\rm P} = K_{\rm P, 0} = 151.35$\,\kms, the expected value (from the values in Table~\ref{tab:param_sys}). 
A departure from 0 for \vv{rad} could be indicative of high-altitude winds in the planet's atmosphere and other dynamic effects (such as planet rotation or a asymmetry between the eastward and westward planet limbs that are probed during the transit) that were not included in the model itself, causing a net blue shift in the observed planetary lines.

\subsubsection{Results}
\label{subsubsec:results_corr}

\begin{figure}[t]
\includegraphics[width=\linewidth]{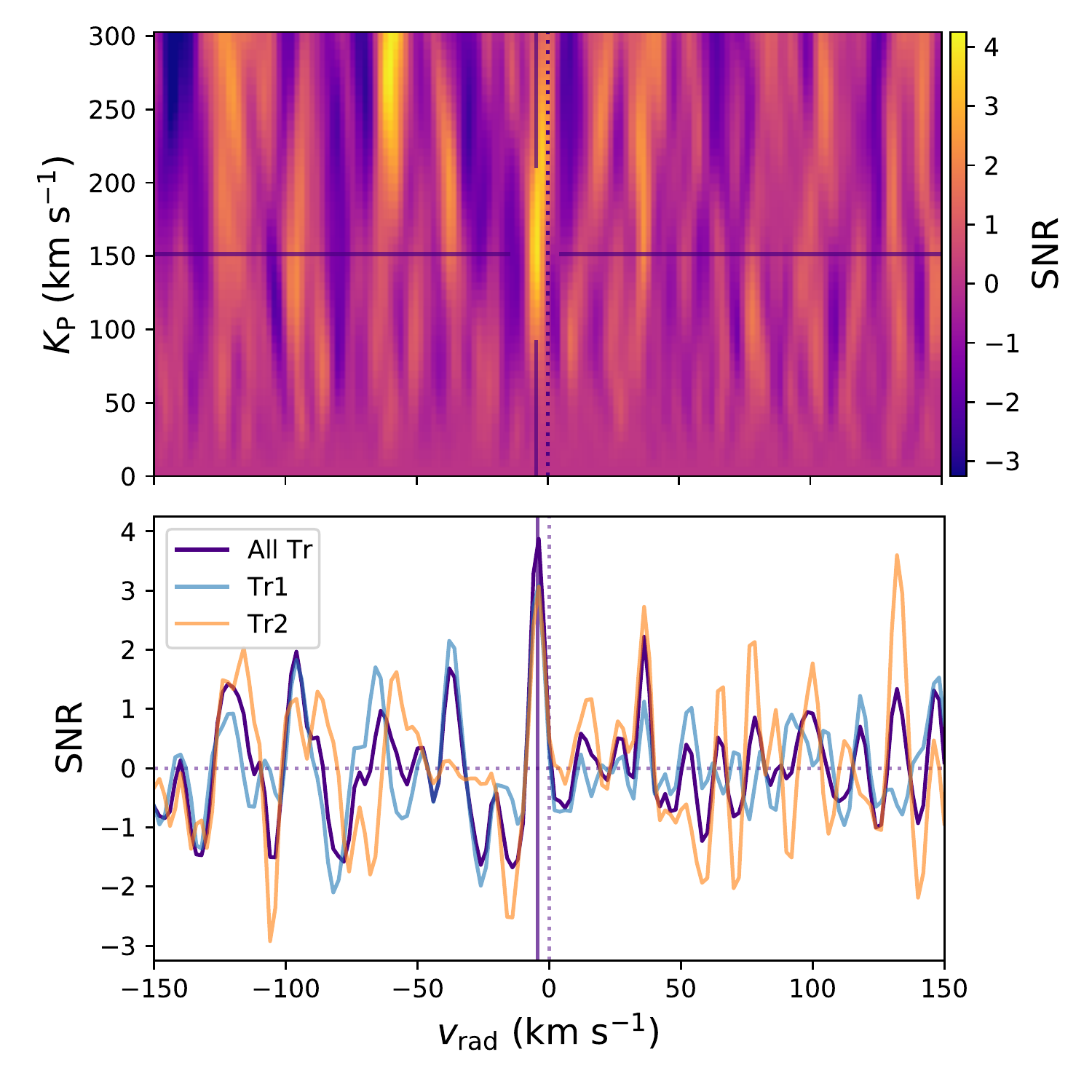}
\caption{ 
[Top panel] Color-coded SNR map of the co-added cross-correlation signal from both transits and the best-fit model ($\log_{10} $VMR[\water]$ = -4.5$, \tp $ = 800$\,K, and \cloud$ = 10^{-0.5}$\,bar) as a function of the zero-phase planet radial velocity and the orbital radial velocity semi-amplitude \kp. 
[Bottom panel] Horizontal cut of SNR map at the known \kp\ of the planet. The peak is found at -4.5\,\kms\ with a SNR of 4.0.
}
\label{fig:CCF}
\end{figure}

\begin{figure}
\centering
\includegraphics[width=\linewidth]{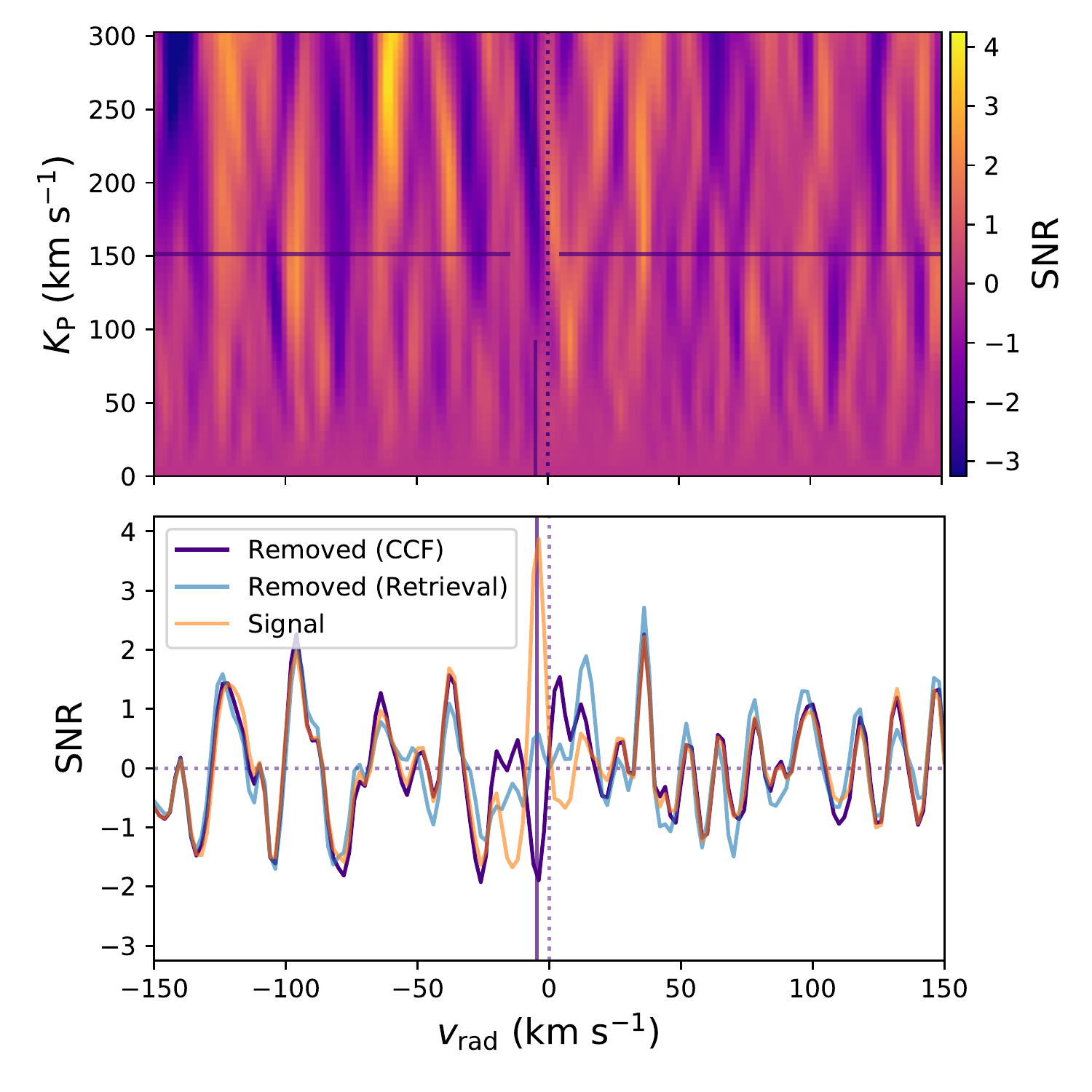}
\caption{[Top panel] Same as Figure~\ref{fig:CCF}, but with the injected negative best-fit model from the CCF grid, with the same color scale.  [Bottom panel] The original signal of the combined transits (orange curve, same as Figure~\ref{fig:CCF}) is reduced considerably when the CCF best-fit model is injected negatively in the data prior to the analysis (indigo curve). When looking at the negative injection of the best-fit model found in the retrieval with the \logl\ (see Section~\ref{subsec:MCMC}), we get a cleaner signal removal (blue curve). In both cases, the systematic noise structures away from the peak are nearly identical as before. }
\label{fig:CCF_no_signal}
\end{figure}

We computed the cross-correlation of a grid of models containing only \water\ as their major constituent. In principle, the molecular or atomic species that are present in both the stellar and planetary atmospheres will introduce signal contamination via the RME when trying to isolate the planetary signal. However, we ignored this effect as water is not expected to be present in the atmosphere of a star like HD~189733.
The explored parameter values on the grid were the following: volume-mixing ratios (VMRs) for water ranging from \vmr\ = -8 to -1.5 with steps of 0.5, equilibrium temperatures \tp\ (input to the Guillot T-P profile) ranging from 300 to 2000\,K in steps of 100\,K and grey cloud deck top pressures of $\log_{10} P_{\rm cloud}$ (bar) from -5 to 2, also in steps of 0.5. 
For this, the \vv{rad} range goes from $-70$ to 70\,\kms, with 2\,\kms\ steps (71 steps in total; roughly the size of a SPIRou pixel, $\simeq 2.3$\,\kms), but \kp\ is fixed at $K_{\rm P, 0}$. 
When combining the two transits, we obtain the maximum CCF SNR grid shown in Figure~\ref{fig:grid_ccf}. The best-fit model has $\log_{10} $VMR[\water]$ = -4.5$, \tp\ $ = 800$\,K and \cloud\ $ = 10^{-0.5}$\,bar, at \vv{rad}$ = -4.5$\,\kms.

To obtain a better estimation of the detection significance of the best fit model, we repeated all calculations for different combinations of \vv{rad}\ and \kp\ (yielding different projected planet radial velocities during the transit sequence). We extended the \vv{rad}\ range from -150 to 150\,\kms\ (still with 2\,\kms\ steps) to get a broader view of the map, while the \kp\ values range from $0$ to $2\,K_{\rm P, 0} \simeq 303$\,\kms, with $\sim 6$\,\kms\ steps (51 steps in total). This gives us the total cross-correlation signal map as a function of \kp\ and \vv{rad}. Again, to produce a SNR detection map, we divided the cross-correlation map by the standard deviation of the full map computed by excluding the region around the peak ($\pm 15$\,\kms\ in \vv{rad}\ and $\pm 70$\,\kms\ in \kp). We did not include the negative \kp\ region, but verified and confirmed that no spurious signal could be found at -\kp$_{, 0}$. 

\begin{figure}
\hspace*{-0.4cm}
\includegraphics[scale=0.38]{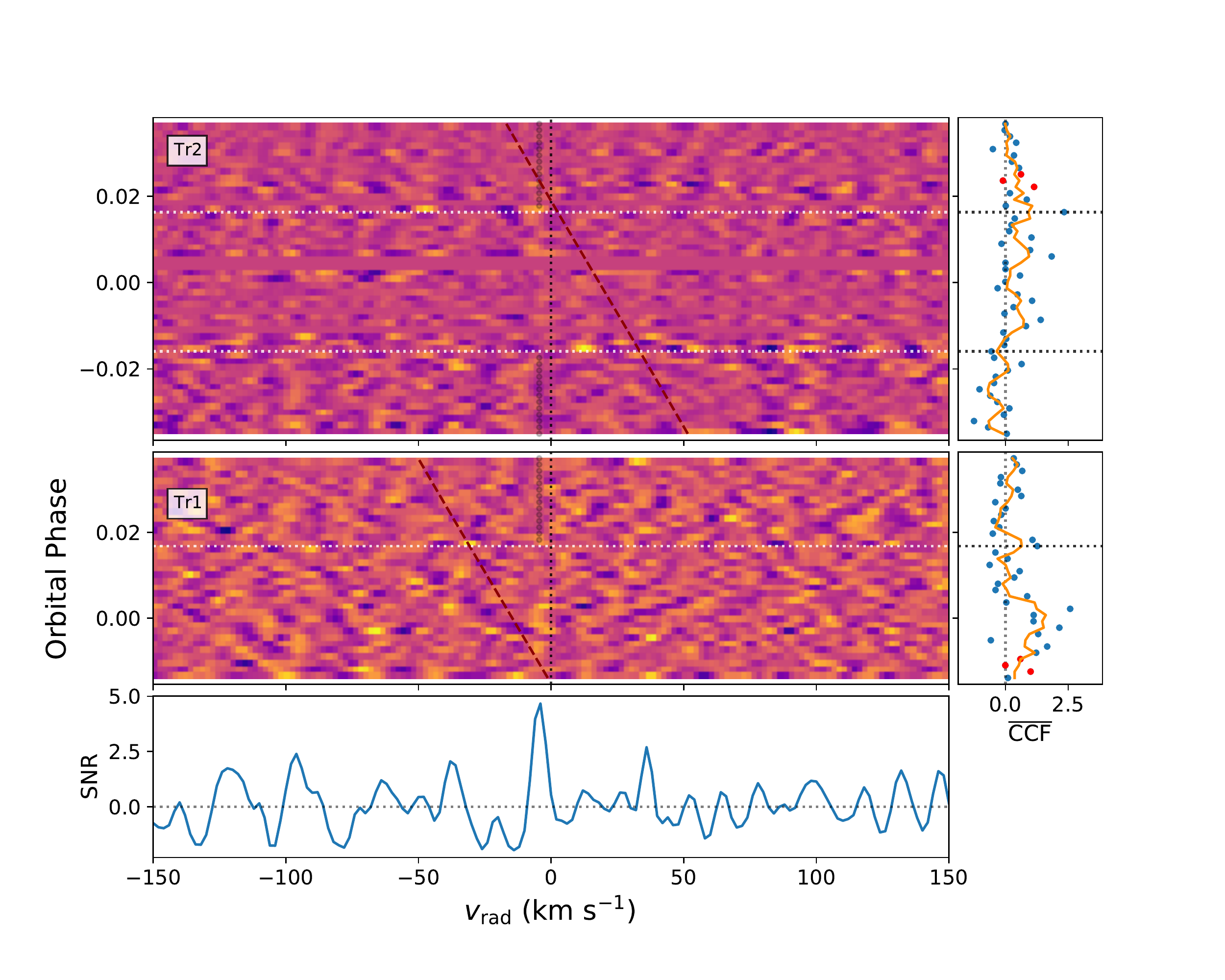}
\caption{[Left] CCF (normalized by the dispersion of the region away from the peak) from of individual spectra as a function of \vv{rad} for Tr2 (top panel) and Tr1 (middle panel), in the planet rest frame. Red dashed lines show the position of the BERV, while whites (left) and black (right) dotted lines shows the ingress and egress positions. The vertical dotted lines show the planet path for \vv{rad} $ = 0$\,\kms\ and \vv{rad} = \vv{peak} (only shown out-of-transit to increase visibility during transit). The bottom panel shows the combined transits CCF SNR curve. 
[Right] Mean CCF for a 3-pixel column bin centered on \vv{peak} (blue points) and the 3-exposures binned signal (orange line). The red dots show where the BERV crosses the peak position by less than 2.3\,\kms. We can see that the detection is not significantly affected by these points.}
\label{fig:CCF_2d}
\end{figure}

The full CCF map of the best-fit model is shown in Figure~\ref{fig:CCF}. We detect \water\ with an SNR $= 4.0$ at \vv{rad}\ $ = -4.5$\,\kms\ (at $K_{\rm P, 0}$). 
We note that the position in \kp\ has large uncertainties ($153^{+57}_{-37}$\,\kms), but these are expected due to the small variation in orbital velocity of the planet during the transit. It is nonetheless still consistent with $K_{\rm P, 0}$. Moving forward, we assume \kp\ is well known and equal to $K_{\rm P, 0}$. Also, the peak is clearly blue-shifted from the planetary rest frame ($v_{\rm rad}=0$). This is discussed in more details in Section~\ref{subsec:wind}. The CCFs of each individual spectrum using the best-fit model are shown in Figure~\ref{fig:CCF_2d}.

When considering only Tr1 or only Tr2, we get a weaker detection: a peak SNR of $ \sim 3.2$ is observed at \vv{rad} $ = -4.7$\,\kms\ for Tr1, while for Tr2, we get a peak SNR of $ \sim 3.1$ at \vv{rad} $ = -4.3$\,\kms. These results, even though they are weaker detections on their own, are consistent within the uncertainties with one another and with the combined transits detection. This further supports the idea that our detected peak is physical as opposed to being a noise artefact. Again, the low SNR from Tr2 could partly be explained by the high variation in exposure SNR (see Figure~\ref{fig:SNR}).

To make sure that the atmospheric parameters that were retrieved by the CCF SNR method are reliable, an injection test was performed. By injecting a negative version of the best-fit model in the data, the original CCF peak should nicely disappear. We did such an injection in the telluric-corrected data using the best-fit model, applying \vv{rad} $ = -4.5$\,\kms, and reapplied all the analysis steps (from Section~\ref{subsec:transpec}). We confirmed that when doing so the CCF peak signal indeed disappears, as can be seen on Figure~\ref{fig:CCF_no_signal}. The same injection test was also verified for the \logl\ approach of the next section (not shown).


As a validation of our adopted approach of applying the data processing steps to the model spectra before calculating the CCF/\logl, we repeated our analysis, but using ``unprocessed" models\footnote{Only a median normalization was applied, but most importantly no PCs were removed to the synthetic sequence.}, and then performed the same negative injection test. When the best-fit models found by the ``unprocessed models"\footnote{The best-fit parameters differed by a factor $10^3$ in VMR[\water], while having similar \tp\ compared to those based on the ``processed models" analysis.} analysis were negatively injected in the data before carrying out the analysis with unprocessed models, the detection peak could not be adequately removed (not shown here). This indicates that applying the data processing to the models yields best-fit parameters that are more representative of what is initially present in the data, and thus that this approach is to be preferred.

\begin{figure}
\centering
\hspace*{-0.5cm}
\includegraphics[scale=0.48]{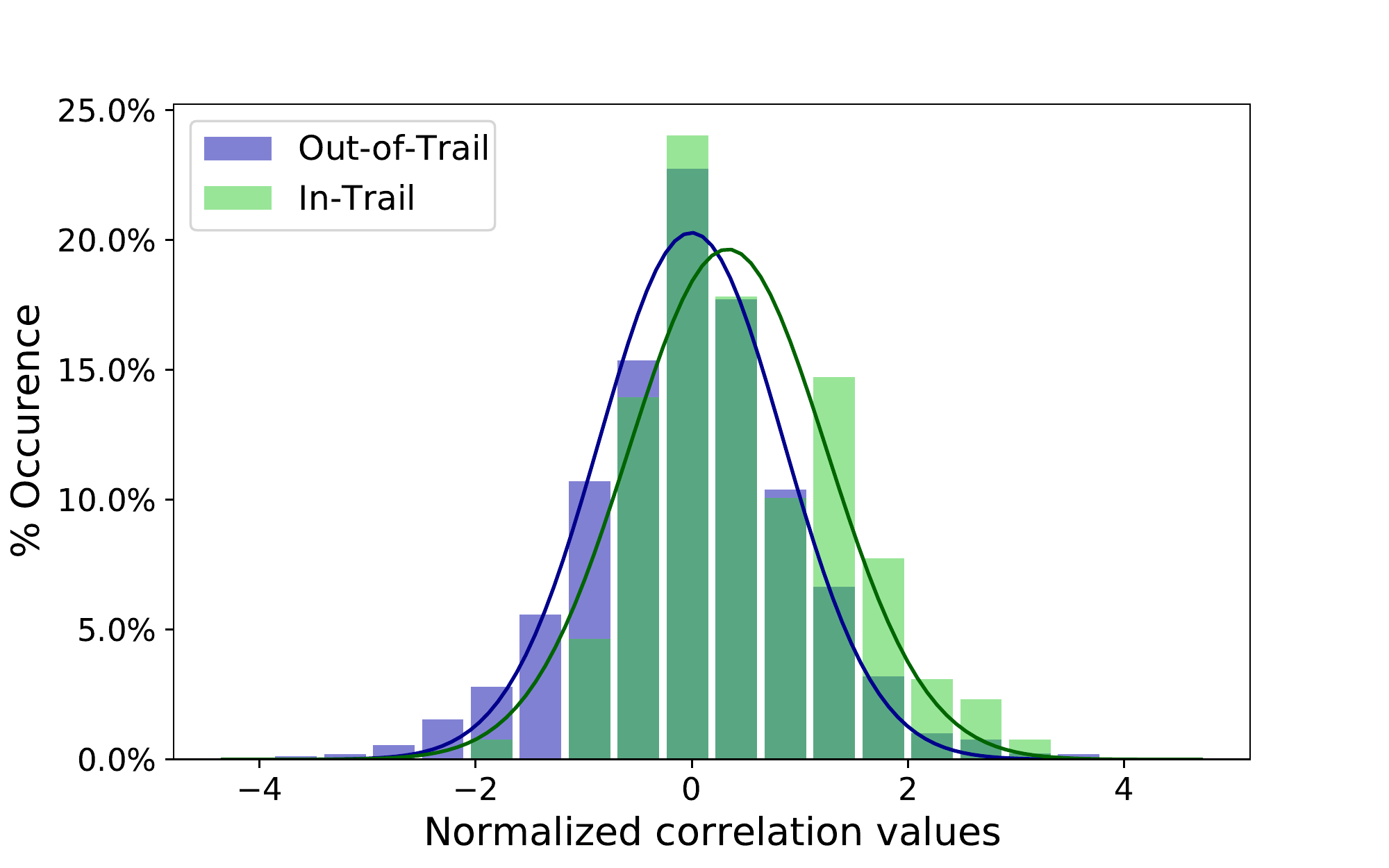}
\caption{Generalized \textit{t}-test results : 
Distributions of cross-correlation values, normalized by the dispersion of the out-of-trail regions, away from (out-of-trail, blue), and near (in-trail, green) the planet radial velocity (with $K_{\rm P, 0}$ and \vv{rad} $ = -4.5$\,\kms), with their associated best-fit Gaussian distributions, each with their corresponding mean and variance (blue and green lines, respectively). A detection of the transmission signal of the planet is expected to shift the in-trail distribution to higher correlation values, and this is what we observe, with the two distributions differing at the $5.9\,\sigma$ level. }
\label{fig:ttest}
\end{figure}

We looked at another statistical metric to differently quantify the significance of our \water\ detection. 
We performed a Student's \textit{t}-test \citep{student1908}, similarly to what was done in other studies before \citep[e.g.][]{Birkby2017, Brogi2018, Cabot2019, Alonso2019_water, Webb2020}. The \textit{t}-test verifies the null hypothesis that two Gaussian distributions have the same mean value. In our case, the two distributions to be compared are drawn from our correlation map. On one hand, we have the in-trail distribution of CCF values, i.e.\ where the planet signal should be, that we took to be correlation values from 3-pixels wide columns centered at \vv{rad} $ = -4.5$\,\kms (as done in \citealt{Birkby2017} and suggested in \citealt{Cabot2019}), and on the other hand, we have the out-of-trail distribution, which includes the CCF values more than 10\,\kms\ away from $v_{\rm P}(t)$, where there should be no planet signal. The \textit{t}-test then evaluates the likelihood that these two samples were drawn from the same distribution.

\begin{figure}
\hspace*{-0.4cm}
\includegraphics[scale=0.50]{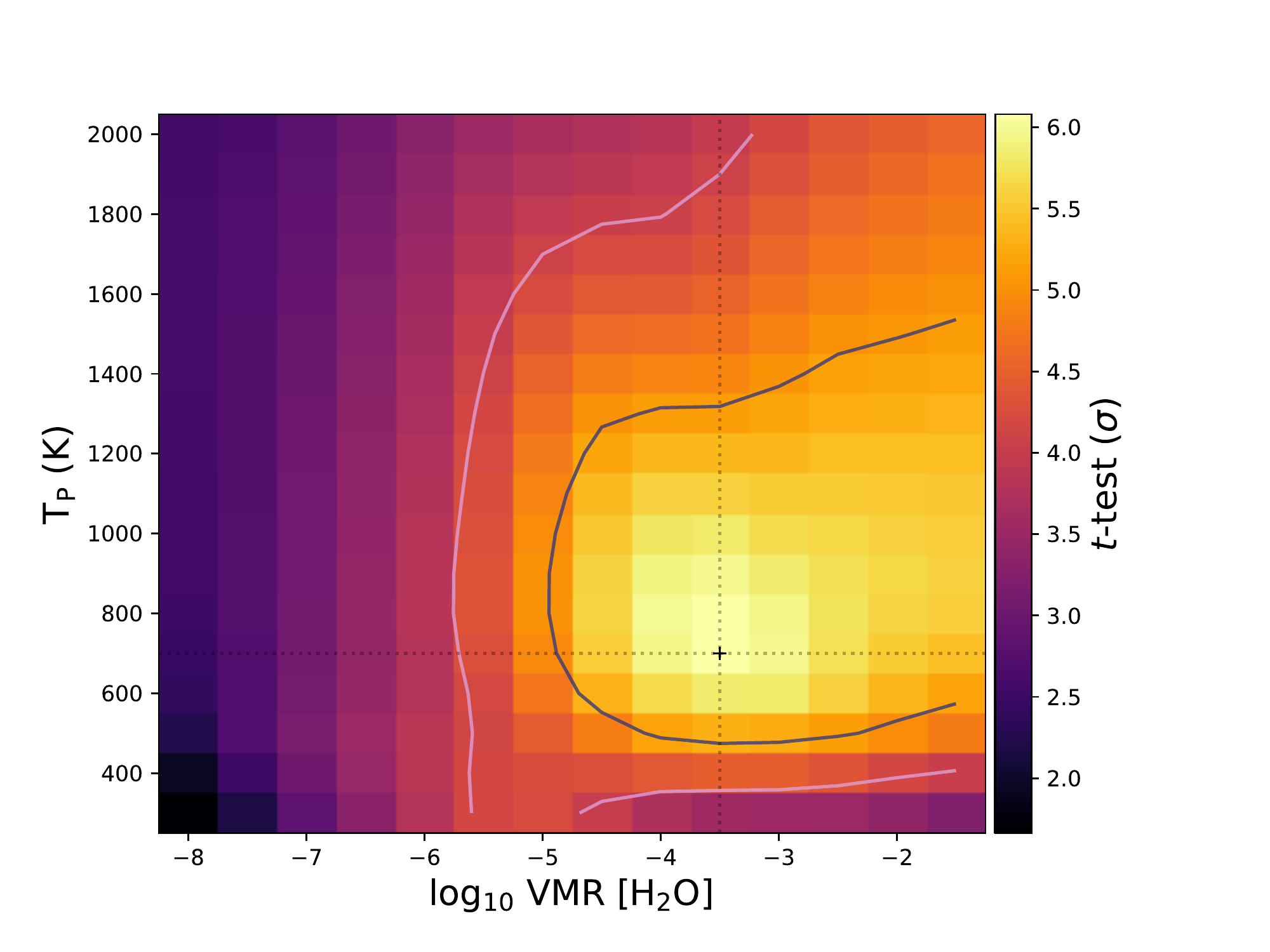}
\caption{Same as Figure~\ref{fig:grid_ccf} but for the $t$-test value. The contour shows where $\sigma$ drops by 1 and 2 from the maximum. A maximum 6.1$\sigma$ is found for $\log_{10} $VMR[\water]$ = -3.5$, \tp $ = 700$\,K, and \cloud$ = 10^{-0.5}$\,bar.}
\label{fig:grid_ttest}
\end{figure}

We computed this test using the in-transit CCFs from Figure~\ref{fig:CCF} with \kp$ = K_{\rm P,0}$ and \vv{rad}$ = $\vv{peak}$ = -4.5$\,\kms. 
The results are shown in Figure~\ref{fig:ttest}, for the combined transits, and indicate that the in-trail distribution is different from the out-of-trail one at the level of $5.9\sigma$, which further supports our detection. As a sanity check, we also compared the in- and out-of-trail distributions for the out-of-transit CCFs. As expected, we found that both distributions/samples had similar mean and variance (different at 0.4\,$\sigma$; not shown here). 
This Student \textit{t}-test is complementary to the CCF SNR method as it uses a different approach to assess uncertainties from the underlying noise. The high \textit{t}-test detection significance gives good confidence in our detection despite the somewhat modest CCF SNR mentioned above. 
As a mean of comparison, we also computed the $t$-test for the same CCF grid described above and the results are show in Figure~\ref{fig:grid_ttest}. The $t$-test yields similar \tp\ and \cloud, but seems to favor overall higher VMRs.

\subsection{Log-Likelihood Mapping Method}
\label{subsec:logL}

In addition to the CCF and \textit{t}-test calculations, we also computed the log-likelihood values for the different models. We used the approach from \cite{Gibson2020} by introducing scaling factors $\alpha$ and $\beta$ for the model and noise, such that the model $m_i$ becomes $\alpha \beta m_i$\footnote{$\alpha$ accounts for any scaling uncertainties of the model and $\beta$ accounts for potential scaling uncertainties of the white noise.}, but we fixed $\alpha=1$ as water is not expected to be present in an extended atmosphere for \hdun\ (see also \citealt{Brogi2019}). By nulling the partial derivative of the standard $\ln \mathcal{L}$ with respect to $\beta$ (i.e., removing the dependency on $\beta$), the log-likelihood function can be written as:

\begin{equation}
\ln \mathcal{L} = -\frac{N}{2} \ln\left[ \frac{1}{N} \left( 
\sum \frac{f_i^2}{\sigma_i^2} + \sum \frac{m_i^2}{\sigma_i^2}
- 2 \sum \frac{f_i m_i}{\sigma_i^2} \right) \right] ,
\label{eq:loglikelihood}
\end{equation}

where the summation is implied over $i$ (both spectral pixels and time). By inspecting this equation, we can see that the first term is a constant (for a given data set, related to the variance of the data) and the last term is related to the CCF from above (eq.~\ref{eq:CCF}). The middle term is related to the variance of the model and introduces the main difference compared to the previous CCF approach. The effect of this term is to reduce the likelihood value for models with higher variances; so for comparable fits to the data, the lower variance (more conservative) model should be preferred.

\begin{figure}
\hspace*{-0.4cm} 
\includegraphics[scale=0.5]{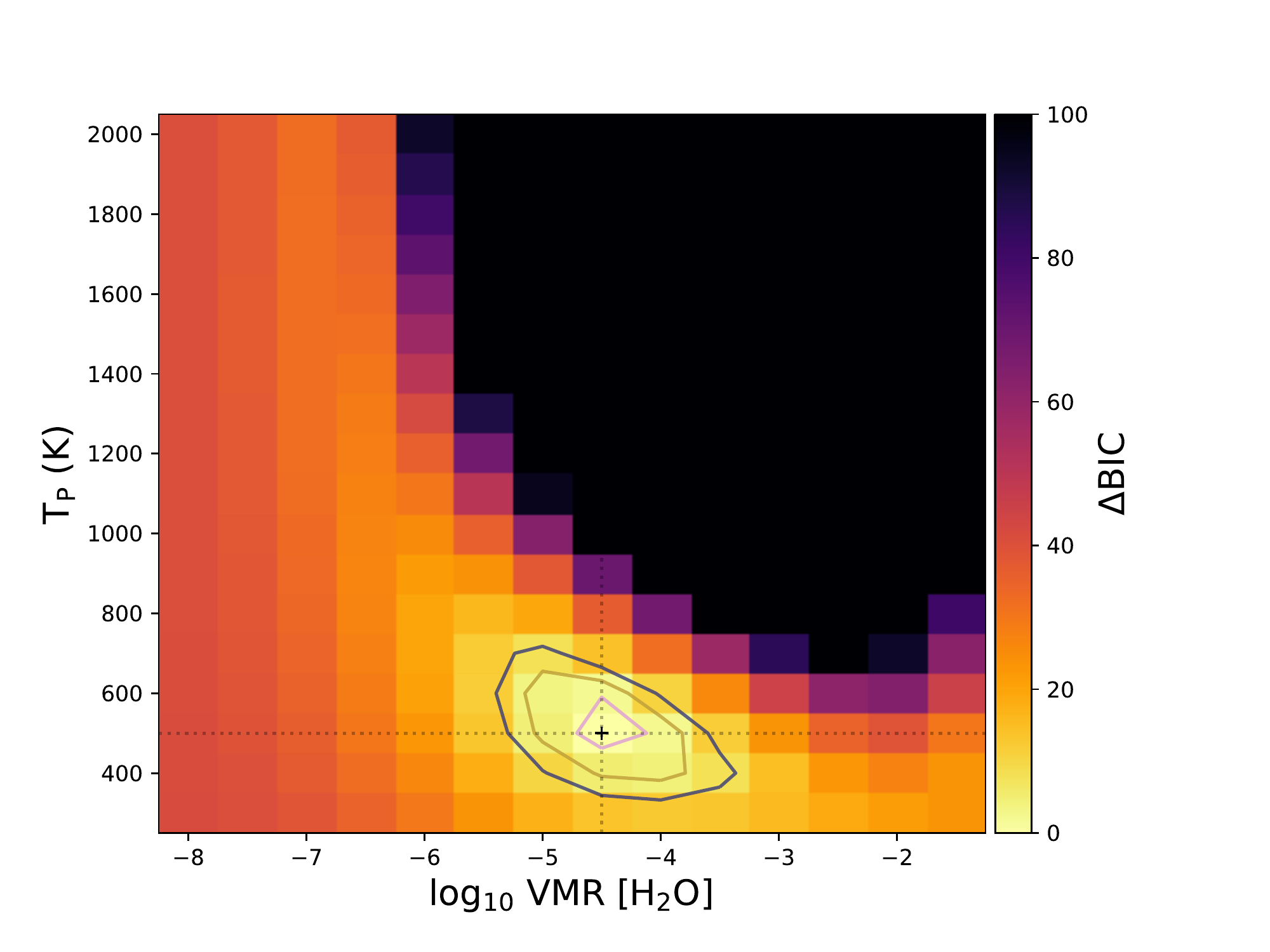}
\caption{\label{fig:grid_logl}
Change in BIC value from the peak, $\Delta$BIC $= 2\,\Delta\,\ln\mathcal{L}$, for the different models in the grid, for a cloud deck top pressure of 1\,bar, at  $K_{\rm P, 0}$ and \vv{rad} $ = v_{\rm peak}$. The contours shown increase by 2, 6, and 10 from the minimum value. For a increase in BIC of 2, 6, and 10, respectively, the best-fit model is typically regarded as being positively, strongly and very strongly favored compared to the other models.} 
\end{figure}

\subsubsection{Results}
\label{subsubsec:results_logl}

For every model in our grid (same as above), we computed the \logl\, for \vv{rad} going from $-70$ to $+70$\,\kms\ and \kp\ $= K_{\rm P, 0}$. The highest \logl\ value occurs around $\log_{10} $VMR[\water] $ = -4.5$, \tp\ $ = 500$\,K, and \cloud $= 1$\,bar, with \vv{rad}$ = -4.7$\,\kms.

Then, to establish how the best-fit model fares compared to the others, 
we looked at the Bayesian Information Criterion (BIC).\footnote{BIC$ = k\ln(n) - 2*\ln\mathcal{L}$; where $k$ is the number of parameters, $n$ is the number of data points and \logl\ is our log-likelihood value for our model for each combination of parameters. For fixed values of $k$ and $n$, the lowest BIC value is related to the highest \logl.} In this formalism, the model with the lowest BIC is preferred (here, taken to be the best-fit model), and the evidence against certain models (with higher BIC) is usually described as ``very strong" when $\Delta$\,BIC $= 2\,\Delta $\logl\ is greater than 10 \citep{Kass1995}. 
Figure~\ref{fig:grid_logl} shows this quantity for the best-fit value of \cloud\ = 1\,bar, as a function of \tp\ and $\log_{10} $VMR[\water]. 
Even though there is a degeneracy between the temperature and the VMR (as well as between the VMR and the cloud deck top pressure, better seen on Figure~\ref{fig:mcmc}), we can still put some constraints on the parameters.   


\subsection{Markov Chain Monte Carlo}
\label{subsec:MCMC}
It quickly becomes time consuming and computationally expensive to compute the full log-likelihood maps for bigger model grids (either in sampling or number of parameters). Instead of computing the log-likelihood for the full grid, we used a Markov Chain Monte Carlo (MCMC) approach that allows us to compute a posterior distribution of each atmospheric parameter and obtain estimates of their uncertainties. The MCMC sampling was done using the \texttt{python} library \texttt{emcee} \citep{emcee}, which implements the affine-invariant ensemble
sampler by \cite{Goodman2010}.

To save time on the generation of models, we applied an N-d linear interpolation\footnote{Using \texttt{SciPy} interpolation module \texttt{RegularGridInterpolator}, which only accepts regular grids, but may have uneven spacing.} over our grid for the parameters VMR[\water], \tp\ (K) and \cloud\ (bar) dimensions. We are aware that these interpolated models are not as accurate,\footnote{As compared to models computed with exact parameters, worst-case interpolated models show differences of at most 62~ppm for the strongest lines, with the biggest errors coming from the cloud top pressure interpolation.} nor are they self-consistent, but they offer a reasonable approximation to derive useful constraints on the atmosphere parameters.

\tabletypesize{\normalsize}
\begin{deluxetable}{lcrc}[tp]
\tablewidth{0pt}
\tablecolumns{4}
\tablecaption{MCMC Retrieval Parameter Priors and Results \label{tab:param_MCMC}}
\tablehead{
\colhead{Parameter}  & \colhead{Uniform Prior} & \colhead{Results} & \colhead{Unit}
}
\startdata
$\log_{10}$ \water & $\mathcal{U}(-8,-1.5)$ & $-4.4^{+0.4}_{-0.4}$ &  \\ 
$T_{\rm P}\,^a$ & $\mathcal{U}(300, 2000)$ & $532^{+46}_{-82}$ & K \\ 
$\log_{10} P_{\rm cloud}$ & $\mathcal{U}(-5, 2)$ & $-0.4^{+1.8}_{-0.3}$ & bar \\
$K_{\rm P}$ & $\mathcal{U}(100, 200)$ & $151^{+10}_{-10}$ & \kms \\ 
$v_{\rm rad}\,^b$ & $\mathcal{U}(-20, 10)$ & $-4.62^{+0.41}_{-0.39}$ & \kms  
\enddata
\tablecomments{ 
\footnotesize
The marginalized parameters from the likelihood analysis with $\pm 1\,\sigma$ error. $^a$ The retrieved \tp\ is the equilibrium temperature input to the Guillot T-P and represents a temperature of $727^{+63}_{-111}$\,K at 1\,bar. $^b$ The uncertainty shown in the last column results from the MCMC analysis only. The uncertainty of $\pm 0.21\,$\kms\ from \vv{sys} must be added to it in quadrature to obtain the total uncertainty on \vv{rad}.
}
\end{deluxetable}

Using the combined transits sequences, we ran 50 walkers for 4500 steps and varied five parameters: $\log_{10}$VMR[\water], \tp, $\log_{10} P_{\rm cloud}$, \kp\ and \vv{rad}. 
We chose to include \kp\ in the retrieval even though this value is well known, to see if it could be retrieved independently and how it would affect \vv{rad}. 

\begin{figure*}
\includegraphics[scale=0.6]{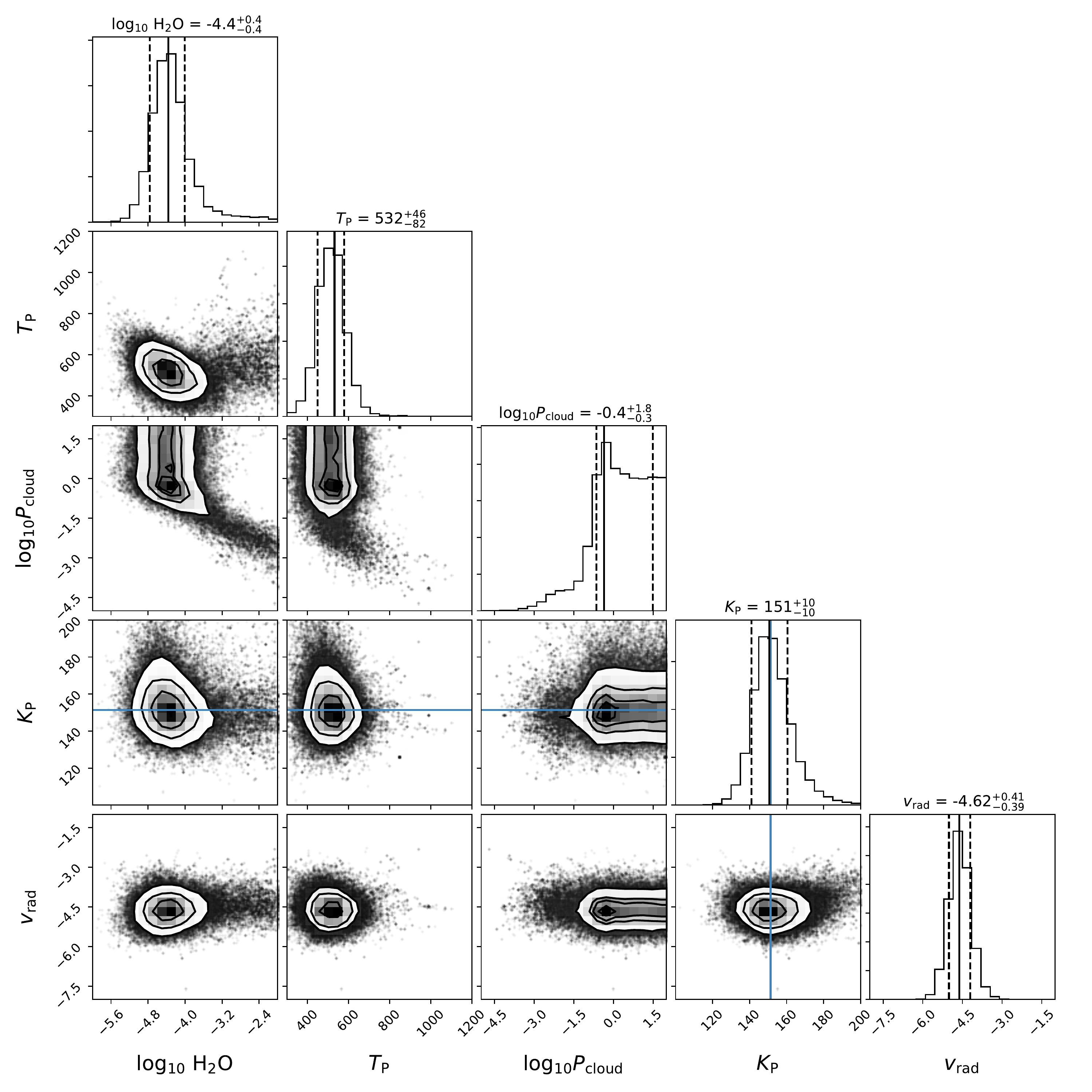}
\caption{  
Posterior probability distributions for the parameters of the MCMC fit on the \hdun\ data using an atmosphere with \water\ and a grey cloud deck as described in Section~\ref{subsec:model}. The blue line shows the position of $K_{\rm P,0}$.}
\label{fig:mcmc}
\vspace*{0.1cm}
\end{figure*}

We are discarding the first 1000 steps of each chain as burn-in\footnote{This is where the chains were overall converged.}. The resulting posterior probability distributions of the five parameters are shown in Figure~\ref{fig:mcmc}. The priors that were used and the resulting marginalized values for each parameter are listed in Table~\ref{tab:param_MCMC}. The 1\,$\sigma$ uncertainties correspond to the range of parameters containing 68\% of the MCMC samples. 

\begin{figure}
\includegraphics[width=\linewidth]{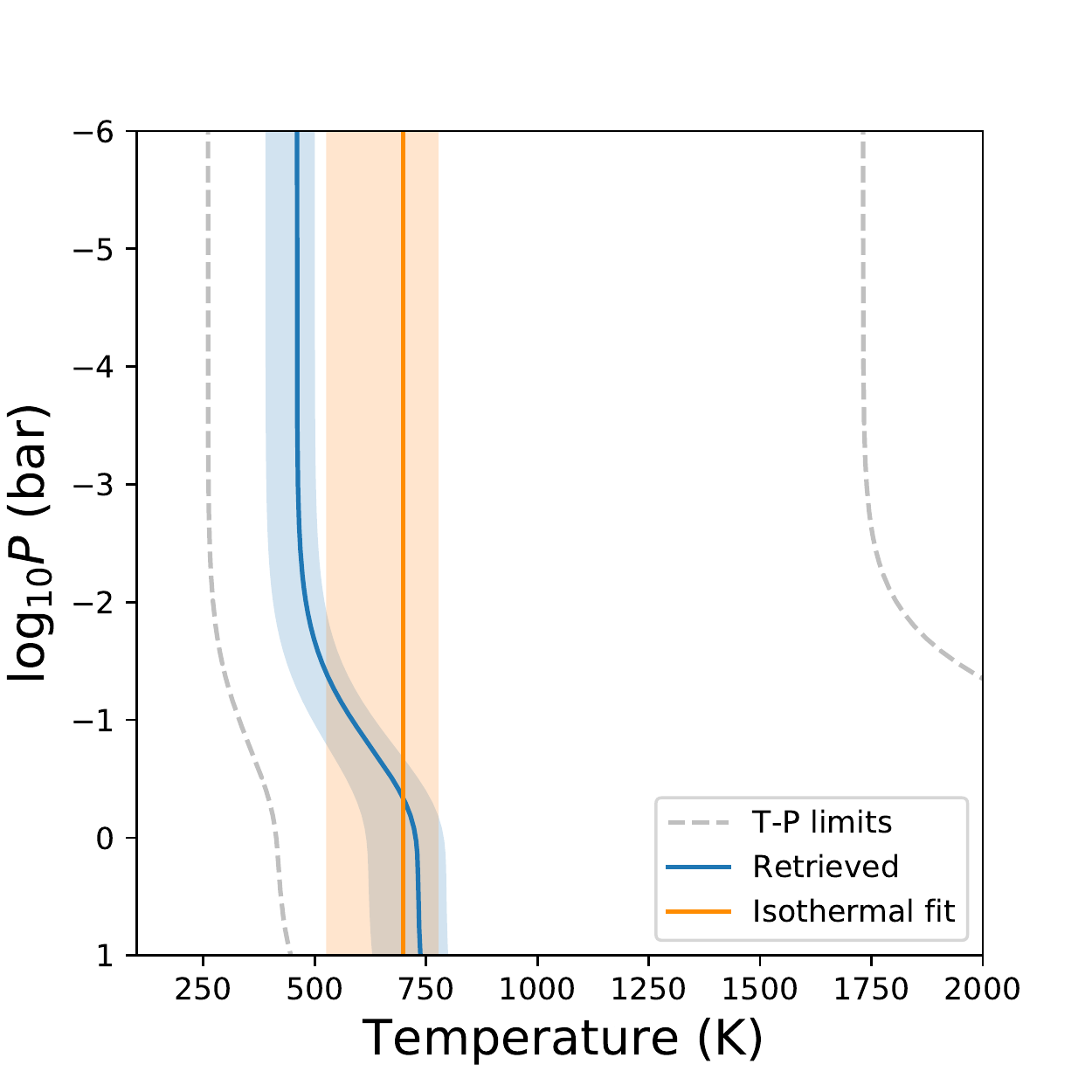}
\caption{  
Retrieved Guillot T-P profile (blue) where the shaded region represents the $1\,\sigma$ uncertainties, compared to the limits of the prior (gray dashed lines). The retrieved profile has a temperature of $727^{+63}_{-111}$\,K at 1\,bar. Overplotted is the retrieved isothermal profile (orange), with \tp $= 698^{+80}_{-172}$\,K, that was obtained from a previous run of the MCMC analysis, that yielded nearly identical values for the other parameters. }
\label{fig:tp-profile}
\vspace*{0.1cm}
\end{figure}

We find relatively good constraints on the parameters. The favored \tp\ and VMR values are close to previous values in the literature (see Section~\ref{subsec:water}), both from low- and high-resolution data. The resulting T-P profile is shown on Figure~\ref{fig:tp-profile}. The cloud top pressure is consistent with relatively deep clouds (at pressures around $\gtrsim 0.2$\,bar) \citep{Mccullough2014, Pinhas2019}. 
However, the apparent degeneracy between the water abundance and cloud top pressure leads to probability at slightly higher cloud altitudes.
This is not a detailed exploration of cloud modeling as in \cite{Barstow2020}, since we did not include the impact of having different cloud fractions (a thorough cloud analysis is beyond the scope of the paper), but our results seem to land somewhere in between their results with and without cloud fractions. 

Nonetheless, the CCF peak disappear when the best-fit from this retrieval is injected negatively in the data, as shown in Figure~\ref{fig:CCF_no_signal}, meaning that it represents well the planetary signal. The cleaner signal removal indicates that this method is probably better at identifying the best model.

\section{Discussion}
\label{sec:Discuss}

Our results demonstrate that water is detected in each of our two SPIRou transmission spectra of \hdun\ (although more marginally in Tr2). Combining both data sets, the standard cross-correlation method yields a detection with SNR = 4.0, based on the dispersion of values in the full CCF \kp\ vs \vv{rad} map, or $5.9\,\sigma$ according to the Student's \textit{t}-test. The peak in the CCF map is located near the expected \kp\ and \vv{rad}, adding confidence to our detection. Similarly, the log-likelihood approach yields a significant detection, with a $\Delta$BIC$>10$ compared to water-less models. A comparison of the results obtained from the model grid for all methods is presented in Table~\ref{tab:param_results}.

\tabletypesize{\normalsize}
\begin{deluxetable*}{lccccccc}
\tablewidth{0pt}
\tablecolumns{8}
\tablecaption{Overview of the results from the model grid for the different methods, using their respective best-fit model. \label{tab:param_results}}

\tablehead{
  &  \multicolumn{3}{c}{Significance $^{\rm a}$} & \multicolumn{4}{c}{Combined Best-fit Model Parameters}
  \\
  \cmidrule(lr{0.25em}){2-4} \cmidrule(l{0.25em}r){5-8}
\colhead{Method}  & \colhead{Tr1} & \colhead{Tr2} & \colhead{Combined}  & \colhead{$v_{\rm rad}$ (\kms)}  & \colhead{$\log_{10}$VMR[\water]} & \colhead{\tp\ (K)} & \colhead{$\log_{10}$\cloud\ (bar)}
}
\startdata
CCF  &  4.4 & 3.0 & 5.0 &  -4.5 & -4.5 & 800 & -0.5  \\ 
$t$-test  &  5.4 & 4.0 & 6.1  &  -4.3 & -3.5 & 700 & -0.5 \\ 
\logl\ $^{\rm b}$ &  24  & 20 & 41  &  -4.7 & -4.5 & 500 & 0
\enddata
\tablecomments{ \footnotesize
$^{\rm a}$ The metrics used are SNR for the CCF method, $\sigma$ for the $t$-test, and $\Delta$BIC for \logl. 
$^{\rm b}$ The $\Delta$BIC value is computed between the best-fit model (i.e. with the largest \logl\ value) and a water-less unconstrained model (i.e. with $\log_{10}$VMR[\water] $= -8$).
}
\end{deluxetable*}


While we obtain a convincing detection using all approaches -- cross-correlation, $t$-test and log-likelihood -- and retrieve similar/consistent best-fit model parameters, we observe some discrepancies for the ``acceptable range" of models from each method (see Figures~\ref{fig:grid_ccf}, \ref{fig:grid_ttest} and \ref{fig:grid_logl}). 
The main discrepancy is that using the CCF (or $t$-test, which is based on the CCF maps), models with a combination of higher temperature and higher \vmr\ seem acceptable (upper-right region on the grid of Figure~\ref{fig:grid_ccf}), while they are strongly disfavored in the log-likelihood analysis. The main difference between the two methods is that the log-likelihood takes into account not only the variance of the observed spectrum, but also the variance of the model, effectively penalizing models with a high variance, and thus preventing higher variance in the model to compensate for a weaker signal in the data. For that reason, we tend to lend more credence to this approach when it comes to constraining the atmosphere parameters. The CCF and $t$-test remain important, however, for confirming a detection of the planet atmosphere.

\subsection{Effects of telluric residuals}

As explained above, we masked spectral regions affected by deep telluric lines before computing the CCF and log-likelihood, where deep was defined here as lines reaching below 30 to 35\% transmission in their core. Applying this mask provided a gain in CCF peak SNR\footnote{We focus this discussion on the CCF since this is one of the most common metric in the literature, but the same general behaviour was observed for the two other metrics, i.e. $t$-test and \logl, in what follows, except when specified.}, even though the spectra were corrected for telluric absorption by \texttt{APERO}. This likely indicates that the telluric correction is not perfect and that a low level of residuals persists. Masking these residuals reduces the number of data points that can be used, but doing so still improves the significance of the detection. As an indication of the magnitude of the effect, if we do not mask deep telluric lines (beyond the masking of saturated lines done by \texttt{APERO}), the peak CCF SNR is decreased by $\sim 2$ when removing a single PC in step 6 from Section~\ref{subsec:transpec} (but decreases by only $\sim 0.6$ when removing three) as compared to the value with additional masking. This implies that there most likely is telluric residual contamination, but that removing three PCs is an effective way to subtract them.

The presence of telluric residuals would also affect the different orders to varying degrees, as the telluric lines are not spread uniformly -- in density and strengths -- across the orders. In our analysis above, we included all spectral orders\footnote{Again, except those heavily filled with saturated telluric lines that are automatically masked by \texttt{APERO}.} in our CCF and log-likelihood calculations, but in some previous studies, the spectral regions most affected by tellurics have sometimes been omitted from the calculations. We tested this to see the effect on the peak CCF SNR, and indeed found that masking some orders could lead to improvements. For instance, when masking 18\% of the orders\footnote{The orders centered at 1.48, from 1.75 to 1.97, plus 2.20 and 2.26\,$\mu$m, which correspond mostly orders between the H and K bands and a few others.}, the peak of the 2D map can increase to 5.3 (from the value of 4.0 when including all orders), while its position does not change much. However, this cut did not improve the $t$-test nor \logl\ values and even led to smaller values. 
Also, for the model grids, the order cut moves the favored models to slightly higher temperatures.  

Nonetheless, we did not adopt this approach for our analysis above, to avoid the risk that the specific choice of orders to omit would introduce a bias in the favored atmosphere model. We preferred to use as much as possible of the spectral range provided by SPIRou, at the cost of a somewhat lower detection SNR. 
Still, the retrieved SNR of 4.0 in CCF for the combined transits is quite smaller than previous detections, e.g. the water detection with an SNR of 6.6 from \cite{Alonso2019_water}, coming from a single transit using CARMENES data, with lower overall SNR, a smaller spectral range, but a slightly better spectral resolution. Our smaller SNR could be explained by multiple reasons. 
First, it could be due to the subjective character of the determination of the SNR with the map standard deviation \citep{Cabot2019}. Variables such as the extent on which the CCF map and its standard deviation are computed, and the \vv{rad} and \kp\ sampling, can greatly change the results (our goes from $\pm 150$\,\kms\ with steps of $2\,$\kms, while theirs goes from $\pm 65$\,\kms\ with steps of $1.3\,$\kms). Computing the SNR from the 2D map or a 1D array can also change the results. The \textit{t}-test is also subject to arbitrarity through the RV sampling that can increase the sample size of both the in- and out-of-trail distributions \citep{Cabot2019}, but is usually preferred to determine detection significance \citep[e.g.,][]{Birkby2017, Brogi2018}.
Second, the lower SNR could also be due to our conservative choice not to remove any spectral orders. 
Third, in our approach the CCF is computed with an injected and reprocessed modeled transit sequence; the reprocessing thus applied to the model can diminish the overall line contrast, which would decrease the CCF (but will also be more representative of the processed signal buried in the observed data). 
Finally, the position of the tellurics relative to the planet signal can affect the results. If the tellurics are not properly corrected/removed, any residuals crossing the planetary path could erroneously increase the retrieved planet signal.



\subsection{Retrieved Atmospheric Parameters}
\label{subsec:water}

Our retrieval results indicate that the probed region of the atmosphere has a temperature that is much lower than the equilibrium temperature ($\tp=531^{+47}_{-81}$\,K, equivalent to $T = 727^{+63}_{-111}$\,K at 1\,bar, while $T_{\rm eq}=1200$\,K), a water abundance that is slightly sub-solar ($\vmr =-4.4^{+0.4}_{-0.4}$, solar abundance would be around $-3$ and $-3.3$ for \tp $ \lesssim$ and $>$ 1200\,K, respectively, \citealt{Madhusudhan2014}, see their Figure~3), and a mostly clear atmosphere (cloud deck at pressure $\gtrsim 0.2$\,bar). 
These results are in line with previous studies at both low and high spectral resolution. For instance, values of \vmr\ between $-3.3$ and -5 for \tp\ between 700--800\,K, or fixed at the equilibrium value, have been reported \citep{Mccullough2014, Madhusudhan2014, Welbanks2019, Pinhas2019}\footnote{Most of those results were obtained using a retrieval with a 6-parameter temperature-pressure (T-P) profile, as in \cite{Madhusudhan2009}.}  at low-resolution; and between $-3$ and $-5$ for models with an atmosphere temperature of around 500\,K (from parameterized, non-isothermal T-P profiles) at high resolution \citep{Birkby2013, Brogi2016, Brogi2018, Alonso2019_water, Brogi2019}. Our results are thus within the range of values found previously, and consistent with sub-solar abundances.



In \cite{Madhusudhan2012_CsurO}, they list two interpretations for low inferred water abundances in hot Jupiters such as \hdun. 
1) The water measurement could be representative of the bulk oxygen abundance of the planet's atmosphere, which would indicate a sub-stellar oxygen abundance, and thus, an overall low planetary atmosphere metallicity. This would be potential evidence for a water-poor formation scenario (with solar relative abundances of the elements, i.e.\ C/O $\sim 0.5$). 
2) For planets with T$\rm _{eq}\gtrsim 1200$\,K, CO becomes a major reservoir for atmospheric oxygen. Therefore, our retrieved sub-stellar water abundance could instead be due to a super-stellar C/O ratio for HD\,189733\,b (in which case more than half of the oxygen would be locked in CO).  This would potentially indicate a formation history dominated by high C/O gas beyond the ice-line, as opposed to solid accretion of oxygen-rich material \citep{Cridland2019_CsurO}. We note, though, that this work focuses solely on the detection of water which prevents us from lifting the degeneracy between these two scenarios. 

When looking at previous strong CO detections and their tendency towards slightly super-solar abundance (at around $\log_{10} [\mathrm{CO}] \simeq -3$; \citealp{deKok2013_SVD, Brogi2016, Cabot2019, Flowers2019, Brogi2019}), combined with our retrieved sub-solar \water\ abundance, this is consistent with high C/O ratio (closer to 1 than solar). This seems to be on trend with other hot Jupiters, such as \hddeux\ \citep{Gandhi2020} and $\tau$~Boo\,b \citep[][; and references therein]{Pelletier2021}.

Recent studies based on 3D atmosphere models have shown that retrieval results based on 1D models, as have been used here, may be biased \citep{Flowers2019, MacDonald2020, Beltz2021}. For instance, \cite{MacDonald2020} analytically showed that a compositional difference between the morning and evening side of the terminator could lead to retrieved 1D uniform temperatures that are many hundreds of degrees colder than the real average terminator temperature. This could possibly explain why, similarly to the previously mentioned results, we retrieved a low \tp\ value. 
They also show that species distributed uniformly around the terminator (such as \water) are biased towards higher abundances. In addition, the omission from our models of CO and other molecules likely to be present in the atmosphere of HD\,189733\,b could also bias the retrieved abundance of water \citep{Brogi2019,MacDonald2020}.
The retrieved parameters in the present work should thus be used with caution - although the magnitude of the aforementioned effects is likely to be small compared to our (relatively large) uncertainties in \tp\ and \vmr.



\subsection{Radial velocity offset}
\label{subsec:wind}

In many previous studies of \hdun, a net blue-shift of the planet atmosphere absorption signal was observed and usually attributed to the presence of large-scale high-altitude day-to-night winds and the presence of eastward jets. Such high-altitude day-to-night winds can be partially accounted for by the $v_{\rm rad}$ term in equation \ref{eq:v_planet} (when they are not already included in the atmosphere models used). Additionally, the planetary rotation/eastward jets should increase the blue shift velocity, due to the potential asymmetries between the morning and evening sides of the terminator. The combination of a hotter and more inflated blue-shifted evening terminator and a colder and less inflated morning red-shifted side would lead towards globally blue-shifted velocities \citep{Flowers2019}.
For \hdun, an overall (limb integrated) shift of around -1.5 to -2\,\kms\ was obtained in several dynamical studies, with both low- and high-resolution \citep{Louden2015, Brogi2016, Brogi2018, Flowers2019}. Larger shifts were also observed with values around $-4$\,\kms\ \citep{Alonso2019_water, Brogi2019, Damiano2019} that are still in agreement with the other results, considering their larger individual uncertainties ($\sim$ 1 -- 2\,\kms).  

Here we measure a net blue-shift of $-4.62^{+0.46}_{-0.44}$\,\kms, which is somewhat larger than what was observed in many previous studies, but still consistent with the results of \citeauthor{Alonso2019_water} and \citeauthor{Damiano2019} within the uncertainties. We observe a similar blue-shift using both the CCF and log-likelihood approaches, and in each individual transit, which suggests that it is a real feature of our data. It was also present and at this value regardless of the analysis parameters that were applied (eg. telluric fraction masking, number of PCs, etc.). 

Our -4.6\,\kms\ measurement is compatible within 1--2\,$\sigma$ with the blueshift of $-5.3^{+1.0}_{-1.4}$\,\kms\ observed for the trailing limb of the planet by \cite{Louden2015}, which could imply that the signal we measure is dominated by the trailing (evening) limb. To investigate this, we checked if we could see a difference in the value of \vv{rad}\ between the ingress and the egress in our data, as was done in \cite{Louden2015, Flowers2019}. We indeed saw a signal that was more red shifted during ingress and more blue shifted during egress, which would support the above explanation, but the significance of this signal is too low to claim it with confidence. In addition, an ingress-egress difference could have biased our measured net \vv{rad}\ as the ingress of both of our transits suffered in some way from technical problems. A small part of the ingress from Tr1 is missing, due to a late start of the observation, while there was a substantial drop in data SNR during the first half of Tr2. This imbalance between the ingress and egress coverage and data quality could have biased our net \vv{rad}\ values toward a value more representative of the egress, i.e. more blue-shifted.

Our larger net blue shift could, perhaps, be explained by the different line list that we used to build our models. However, when computing the CCF or \logl\ with the best-fit model built with the HITEMP 2010 water line list (instead of Exomol), we get a virtually identical shift, but with a slightly smaller SNR. Also, the cross-correlation between two models with identical parameters but using two different water line lists (i.e.\ HITEMP and Exomol) shows a relative shift of only $\sim 0.023$\,\kms.  

Taken at face value, a net blue-shift as high as we have measured, if coming from large-scale winds, would be hard to explain by current GCMs, based on the modeled CCF curves from \cite{Flowers2019} for different rotation speed. Including the contribution of the planet rotation can lead to higher values of net blue-shift, but for a synchronous rotation (period of 2.2\,days, consistent with the results of \cite{Flowers2019}) this effect would not be enough: a CCF peak value between -2 and -1\,\kms\ would be expected (see their Figure~12).
A planet rotation period of $\lesssim 1.30$\,days (faster than the synchronous case, yielding an equatorial rotational velocity of $\gtrsim 4.85$\,\kms\ and a maximum wind speed of $2.76$\,\kms) could lead to a net blue-shift closer to $-5\,$\kms\ when considering the rotation and winds together (last row from Figure 12 in \citealt{Flowers2019}). 
Such a fast rotation (faster than synchronous) should lead to a significant rotational broadening. Using the CCF approach with models that were rotationally broadened in a simple manner (Gaussian broadening), the narrowest profiles (no rotation) were always favored, as expected and pointed out in \cite{Brogi2016}. On the other hand, using the log-likelihood approach with the same broadening yielded rotation speed of \vv{rot} $ = 2.0^{+1.5}_{-2.0}$\,\kms which is broadly consistent with synchronous rotation (equatorial \vv{rot} $ = 2.849\,$\kms; \citealt{Flowers2019}), but higher rotation speeds were not excluded by the \logl. An overall smaller net blue shift is observed and decreases with higher rotation speed, but that is still too high to be explained from synchronous rotation. Nonetheless, this seems to indicate that rotational information is present in the data and could partially explain the high blue shift. 




Finally, errors in the absolute SPIRou wavelength calibration are not expected to be larger than a few 0.1\,\kms\ and are thereby not likely to be the cause of the high blue shift that we report.  

\section{Concluding remarks}
\label{sec:conclusion}

We presented one of the first exoplanet atmosphere characterization data sets obtained with the SPIRou high-resolution near-infrared spectrograph recently mounted on the 3.6\,m CFHT within the frame work of the Large Observing Program called the SPIRou Legacy Survey totalling 300 CFHT nights, and demonstrated that this instrument is capable of characterizing exoplanet atmospheres via transmission spectroscopy.

Our analysis of the data revealed a detection of water in the atmosphere of \hdun\ at an SNR of 4.0 using the standard CCF method, and $5.9\sigma$ from a Student's \textit{t}-test, and that a good atmosphere model selection can be achieved using the more recent log-likelihood mapping methods from \cite{Brogi2019} and \cite{Gibson2020}. This allowed us to put constraints on the temperature, water volume mixing ratio, and grey cloud deck level of the planet atmosphere. 
Our results favour temperatures that are significantly lower than $T_{\rm eq}$ (\tp\ $\simeq 532\,$K, equivalent to $T = 727^{+63}_{-111}$\,K at 1\,bar), a sub-solar$\vmr \simeq -4.4$, and a grey cloud deck at pressures of $>0.2$\,bar - all within the ranges of values found previously in the literature. 
Moreover, the absorption signal of the planetary atmosphere is detected with a substantial blue-shift of $-4.62^{+0.46}_{-0.44}$\,\kms - a value 0.6--3\,\kms\ bluer than previous literature results. The cause of this difference remains unclear, but it seems unlikely that such a high blue-shift can arise only from winds in the planet's atmosphere. The planet's rotation and/or a signal dominated by the trailing limb of the planet could be at play here. 

This analysis focused mainly on the detection of water, but much more remains to be done with these data. Given the spectral coverage of SPIRou, which extends to the end of the $K$ band, it should be possible to probe for the presence of carbon monoxide in the atmosphere of \hdun. This analysis will require an explicit treatment of the Rossiter-McLaughlin effect, as CO is also present in the stellar spectrum which will introduce time-varying CO features during transit. The SPIRou spectral range also covers the helium metastable triplet at 1.083\,$\mu$m, which probes the extended atmosphere. This absorption was previously detected for \hdun\ \citep{Salz2018, Guilluy2020} using CARMENES and GIARPS (GIANO + HARPS) data, respectively, but is also seen in SPIRou \citep{Donati2020} and it will be interesting to further compare these results.

This work relied on 1D atmosphere models with an isothermal temperature structure. In the future, it will be interesting to analyze the data using more complex models, in particular; 3D models that can capture variations in composition or conditions across the planet limb, as well as global atmosphere dynamics. A free spectral retrieval, rather than a retrieval constrained on a model grid as done here, would also be beneficial.

Finally, we point out that the throughput of SPIRou in the $Y$ and $J$ band was recently increased by a factor of 2 and 1.6, respectively, following the replacement of its set of rhomboid prisms (see \citealt{Donati2020}). This should improve the overall quality of future data compared with what was presented here, and should consequently enable better characterization of planetary atmospheres. In particular, the significant increase in throughput in the blue should greatly aid probes of the 1.083\,$\mu$m metastable helium line used to study the extended atmosphere. As improvements in the data reduction software of SPIRou are continually being made, including better correction of telluric absorption, the overall capabilities of SPIRou for characterizing exoplanet atmospheres are also expected to improve.
\\
\\
\emph{Acknowledgements}

\small
The authors acknowledge financial support for this research from the Natural Science and Engineering Research Council of Canada (NSERC), the Institute for Research on Exoplanets (iREx) and the University of Montreal (UdeM).
These results are based on observations obtained at the Canada-France-Hawaii Telescope (CFHT) which is operated from the summit of Maunakea by the National Research Council of Canada, the Institut National des Sciences de l'Univers of the Centre National de la Recherche Scientifique of France, and the University of Hawaii. The observations at the Canada-France-Hawaii Telescope were performed with care and respect from the summit of Maunakea which is a significant cultural and historic site. 
We thank the reviewer for her/his careful reading and feedback that helped us improve the quality of this manuscript.
SP and ADB acknowledge funding from the Technologies for Exo-Planetary Science (TEPS) CREATE program.
JFD acknowledges funding from the European Research Council under the H2020 research \& innovation programme (grant \#740651 NewWorlds). 
VB acknowledges that this work has been carried out in the frame of the National Centre for Competence in Research “PlanetS” supported by the Swiss National Science Foundation (SNSF). This project has received funding from the European Research Council (ERC) under the European Union's Horizon 2020 research and innovation programme (project {\sc Four Aces} grant agreement No 724427; project {\sc Spice Dune}; grant agreement No 947634).
XD, GH and EM acknowledge funding from the French National Research Agency (ANR) under contract number ANR-18-CE31-0019 (SPlaSH). 
XD acknowledges funding in the framework of the Investissements d'Avenir program (ANR-15-IDEX-02), through the funding of the ”Origin of Life” project of the Univ. Grenoble-Alpes. 
BK acknowledges funding from the European Research Council under the European Union's Horizon 2020 research and innovation programme (grant agreement \#865624, GPRV). 
J.H.C.M. is supported in the form of a work contract funded by Funda\c{c}\~ao para a Ci\^encia e a Tecnologia (FCT) with the reference DL 57/2016/CP1364/CT0007; and also supported from FCT through national funds and by FEDER-Fundo Europeu de Desenvolvimento Regional through COMPETE2020-Programa Operacional Competitividade e Internacionalização for these grants UIDB/04434/2020 \& UIDP/04434/2020, PTDC/FIS-AST/32113/2017 \& POCI-01-0145-FEDER-032113, PTDC/FIS-AST/28953/2017 \& POCI-01-0145-FEDER-028953, PTDC/FIS-AST/29942/2017.
NS acknowledges funding FCT - Funda\c{c}\~ao para a Ci\^encia e a Tecnologia through national funds and by FEDER through COMPETE2020 - Programa Operacional Competitividade e Internacionaliza\c{c}\~ao by these grants: UID/FIS/04434/2019; UIDB/04434/2020; UIDP/04434/2020; PTDC/FIS-AST/32113/2017 \& POCI-01-0145-FEDER-032113; PTDC/FIS-AST/28953/2017 \& POCI-01-0145-FEDER-028953; PTDC/FIS-AST/28987/2017 \& POCI-01-0145-FEDER-028987. 


\bibliographystyle{apj}
\bibliography{ref, references}


\end{document}